\documentclass{article}
\usepackage{amsmath}
\usepackage{amssymb}
\usepackage{epubtk}
\usepackage{graphicx}
\usepackage{mathbbol}
\usepackage{ifpdf}
\usepackage{psfrag}
\usepackage{epstopdf}
\usepackage{enumerate}
\usepackage{color}

\showlistoftablesfalse

\newcommand{\mnote}[1]
{\protect{\stepcounter{mnotecount}}$^{\mbox{\footnotesize $%
\!\!\!\!\!\!\,\bullet$\themnotecount}}$ \marginpar{
\raggedright\tiny\em $\!\!\!\!\!\!\,\bullet$\themnotecount: #1} }

\newcommand{\kk}[1]{}

\newcommand{\ohyp}{\,\,\overline{\!\!\hyp}}
\newcommand{\pohyp}{\partial\ohyp}

\newcommand{\regular}{$I^+$--regular}

\newcommand{\llambda}{\lambda}

\newcommand{\eean}{\nonumber\end{eqnarray}}

\newcommand{\Mtext}{\Sext}

\newcommand{\mcMext}{\Mext}

\newcommand{\mcA}{\mycal A}

\def\K0{\phi^{K_0}}

\def\X.{\phi^{X}\cdot}

{\catcode `\@=11 \global\let\AddToReset=\@addtoreset}
\AddToReset{equation}{section}

\newcommand{\fourg}{{\mathfrak g }}

\newcommand{\mcN}{{\mycal N}}

\newcommand{\nopcite}[1]{}

  {
   }

\newcommand{\const}{\mathrm{const}}

\newcommand{\mcE}{{\mycal E}}

\newcommand{\mcW}{{\mycal W}}

\newcommand{\nablash}{\nabla{\kern -.75 em
     \raise 1.5 true pt\hbox{{\bf/}}}\kern +.1 em}
\newcommand{\Deltash}{\Delta{\kern -.69 em
     \raise .2 true pt\hbox{{\bf/}}}\kern +.1 em}
\newcommand{\Rslash}{R{\kern -.60 em
     \raise 1.5 true pt\hbox{{\bf/}}}\kern +.1 em}

\newcommand{\hyp}{{\mycal S}}

\newcommand{\threeg}{\gamma}

\newcommand{\mcM}{{\mycal M}}

\newcommand{\mcH}{{\mycal H}}

\newcommand{\bea}{\begin{eqnarray}}
\newcommand{\beaa}{\begin{eqnarray*}}
\newcommand{\bean}{\begin{eqnarray}\nonumber}

\newcommand{\bel}[1]{\begin{equation}\label{#1}}
\newcommand{\beal}[1]{\begin{eqnarray}\label{#1}}
\newcommand{\beadl}[1]{\begin{deqarr}\label{#1}}
\newcommand{\eeadl}[1]{\arrlabel{#1}\end{deqarr}}
\newcommand{\eeal}[1]{\label{#1}\end{eqnarray}}
\newcommand{\eead}[1]{\end{deqarr}}
\newcommand{\eea}{\end{eqnarray}}
\newcommand{\eeaa}{\end{eqnarray*}}

\newcommand{\be}{\begin{equation}}
\newcommand{\ee}{\end{equation}}

\newcommand{\ext}{{\mathrm{ext}}}

\newcommand{\eq}[1]{\eqref{#1}}
\newtheorem{defi}{\sc Coco\rm}[section]
\newtheorem{theorem}[defi]{\sc Theorem\rm}
\newtheorem{Theorem}[defi]{\sc Theorem\rm}

\newtheorem{Conjecture}[defi]{\sc Conjecture\rm}

\newtheorem{Definition}[defi]{\sc Definition\rm}

\newtheorem{Proposition}[defi]{\sc Proposition\rm}

\def \R {\Reel}

\newcounter{mnotecount}[section]

\renewcommand{\themnotecount}{\thesection.\arabic{mnotecount}}

\newcommand{\ednote}[1]{}

\newcommand{\opp}[1]{}

\newcommand{\Sext}{\hyp_{\mathrm{ext}}}
\newcommand{\Mext}{\mcM_{\mathrm{ext}}}

\newcommand{\doc}{\langle\langle \mcMext\rangle\rangle}

\newcommand{\mcB}{{\mycal B}}

\def\emph#1{{\it #1}}
\def\textbf#1{{\bf #1}}

\def\R{{\mathbb R}}

\def\K{{\bf K}}

\def\T{{\Bbb T}}

\newcommand{\changedX}{K}

\def\2{{\overline 2}}

\newcommand{\beqa}{\begin{eqnarray}}
\newcommand{\eeqa}{\end{eqnarray}}

\DeclareFontFamily{OT1}{rsfs}{} \DeclareFontShape{OT1}{rsfs}{m}{n}{
<-7> rsfs5 <7-10> rsfs7 <10-> rsfs10}{}
\DeclareMathAlphabet{\mycal}{OT1}{rsfs}{m}{n}

\newcommand{\sprod}[2]{\,g(#1 ,#2)}
\newcommand{\barsprod}[2]{\langle #1\, , \,#2 \rangle}

\newcommand{\Trace}[1]{\hat{\mbox{tr}} \left(#1 \right)}
\newcommand{\trMH}[1]{\mbox{Trace} \left(#1 \right)}

\newcommand{\tDelta}{\Delta_\delta}

\def\bbbr{\R}
\def\D{\mathrm{d}}
\def\rmD{\mathrm{D}}
\newcommand{\gtiltens}{\tilde{\mbox{\boldmath $g$}}}

\newcommand{\fourast}{\ast}
\renewcommand{\fourg}{{\bf g}}
\newcommand{\fourgx}{\fourg}
\newcommand{\targetg}{\mbox{\boldmath $G$}}

\newcommand{\fourF}{F}
\newcommand{\barF}{\bar{F}}
\newcommand{\barA}{\bar{A}}
\newcommand{\barg}{\bar{\mbox{\boldmath $g$}}}
\newcommand{\barast}{\bar{\ast}}
\newcommand{\barR}{\bar{R}}
\newcommand{\barD}{\bar{\mathrm{D}}}
\newcommand{\barlap}{\bar{\Delta}}
\newcommand{\erpot}{{\mathrm{E}}}
\newcommand{\epspot}{{\varepsilon}}

\renewcommand{\mcM}{M}
\renewcommand{\mcN}{H}

\renewcommand{\changedX}{k}
\newcommand{\KV}{\xi}   
\newcommand{\diffk}{d}   
\newcommand{\Ldim}{N}   
\newcommand{\Compact}{Q}   

\newcommand{\mcrealN}{{\mycal N}}

\begin{document} 

\title{Stationary Black Holes: Uniqueness and Beyond}

\author{%
\epubtkAuthorData{Piotr T.\ Chru\'sciel}
                   {University of Vienna}
                   {piotr.chrusciel@univie.ac.at}
                   {http://homepage.univie.ac.at/piotr.chrusciel}
\and
\epubtkAuthorData{Jo\~ao Lopes Costa}
                 {Instituto Universit\'ario de Lisboa (ISCTE-IUL),
                  Lisboa, Portugal
                  \\
                  Centro de An\'alise Matem\'atica, Geometria e Sistemas Din\^amicos, \\
                  Instituto Superior T\'ecnico, Universidade T\'ecnica de Lisboa, Portugal}
                 {jlca@iscte.pt}
                 {}
\and
\epubtkAuthorData{Markus Heusler}
                 {ITP, University of Zurich, CH-8057 Zurich\\
                   (at the time of writing original 1998 version)}
                 {markus.heusler@rsnag.ch}
                 {}
}

\date{}
\maketitle

\begin{abstract}
The spectrum of known black-hole solutions to the stationary Einstein
equations has been steadily increasing, sometimes in unexpected
ways. In particular, it has turned out that not all
black-hole--equilibrium configurations are characterized by their
mass, angular momentum and global charges. Moreover, the high degree
of symmetry displayed by vacuum and electro-vacuum black-hole
spacetimes ceases to exist in self-gravitating non-linear field
theories. This text aims to review some developments in the subject
and to discuss them in light of the uniqueness theorem for the
Einstein--Maxwell system.
\end{abstract}

\epubtkKeywords{black holes, self-gravitating classical fields,
uniqueness theorems}

\newpage

\epubtkUpdate
    [Id=A,
     ApprovedBy=subjecteditor,
     AcceptDate={29 March 2012},
     PublishDate={14 May 2012},
     Type=major]{%
Major update of the original version by Markus Heusler from 1998. Piotr
T.\ Chru\'sciel and Jo\~ao Lopes Costa succeeded to this review's authorship.
Significantly restructured and updated all sections; changes are too
numerous to be usefully described here. The number of references
increased from 186 to 329.
}

\newpage

\section{Introduction}
\label{sec-INT}

\subsection{General remarks}
\label{subsec-INT-GEN}

Our conception of black holes has experienced several dramatic changes
during the last two hundred years: While the ``dark stars'' of
Michell~\cite{M1784} and Laplace~\cite{L1796} were merely regarded as
peculiarities of Newton's law of gravity and his corpuscular theory of
light, black holes are nowadays widely believed to exist in our
universe (for a review on the evolution of the subject the reader is
referred to Israel's comprehensive account~\cite{Israel:bhreview}; see
also~\cite{CMSciama, Carter:1997im}). Although the observations are
necessarily indirect, the evidence for both stellar and galactic black
holes has become compelling~\cite{MR96Chandra, McClintockRemillard,
  Narayan, NGM, Mueller:2007rz, McClintock:2004ji}. There seems to be
consensus~\cite{ReesSupermassive, KorGeb, MerFer, NarayanMahadevan}
that the two most convincing supermassive black-hole candidates are
the galactic nuclei of NGC~4258 and of our own Milky
Way~\cite{GenzelEisenhauerGillessen}.

The theory of black holes was initiated by the pioneering work of
Chandrasekhar~\cite{SC31b, SC31a} in the early 1930s. (However, the
geometry of the Schwarzschild solution~\cite{KS16b, KS16a} was
misunderstood for almost half a century; the misconception of the
``Schwarzschild singularity'' was retained until the late 1950s.)
Computing the Chandrasekhar limit for neutron stars~\cite{BZ34},
Oppenheimer and Snyder~\cite{OS39}, and Oppenheimer and
Volkoff~\cite{OV39} were able to demonstrate that black holes present
the ultimate fate of sufficiently-massive stars. Modern black-hole
physics started with the advent of \textit{relativistic} astrophysics,
in particular with the discovery of pulsars in 1967.

One of the most intriguing outcomes of the \textit{mathematical}
theory of black holes is the uniqueness theorem, applying to a class
of stationary solutions of the Einstein--Maxwell equations. Strikingly
enough, its consequences can be made into a test of general
relativity~\cite{SadeghianWill}. The assertion, that all
(four-dimensional) electrovacuum black-hole spacetimes are
characterized by their mass, angular momentum and electric charge, is
strangely reminiscent of the fact that a statistical system in thermal
equilibrium is described by a small set of state variables as well,
whereas considerably more information is required to understand its
dynamical behavior. The similarity is reinforced by the
black-hole--mass-variation formula~\cite{BCH73laws} and the
area-increase theorem~\cite{HE73LSS, ChDGH}, which are analogous to the
corresponding laws of ordinary thermodynamics. These mathematical
relationships are given physical significance by the observation that
the temperature of the black body spectrum of the Hawking
radiation~\cite{SH75PC} is equal to the surface gravity of the black
hole. There has been steady interest in the relationship between the
laws of black hole mechanics and the laws of thermodynamics. In
particular, computations within string theory seem to offer a
promising interpretation of black-hole entropy~\cite{GH96SC}. The
reader interested in the thermodynamic properties of black holes is
referred to the review by Wald~\cite{BW96SC}.

There has been substantial progress towards a proof of the celebrated
uniqueness theorem, conjectured by Israel, Penrose and Wheeler in the
late sixties~\cite{ChCo, ChNguyen, Costaelvac} during the last four
decades (see, e.g.,~\cite{PC94DG} and~\cite{PC96HPA} for previous
reviews). Some open gaps, notably the electrovacuum staticity
theorem~\cite{SW92stat, SW93stat} and the topology
theorems~\cite{Gallo95top, Gallo96top, CW94TOP}, have been closed
(see~\cite{PC96HPA, CGS, ChHighDim} for related new results). Early
on, the theorem led to the expectation that the stationary--black-hole
solutions of \textit{other} self-gravitating matter fields might also
be parameterized by their mass, angular momentum and a set of charges
(generalized no-hair conjecture). However, ever since Bartnik and
McKinnon discovered the first self-gravitating Yang--Mills
\textit{soliton} in 1988~\cite{BK88soliton}, a variety of new
\textit{black hole} configurations have been found, which violate the
\emph{generalized no-hair conjecture}, that \emph{suitably regular
  black-hole spacetimes are classified by a finite set of
  asymptotically-defined global charges}. These solutions include
non-Abelian black holes~\cite{VG89firsthair, KMuA90firsthair,
  PB90firsthair}, as well as black holes with
Skyrme~\cite{DHS91Skyrme, HSZ93Skyrme}, Higgs~\cite{BFM92monop,
  OliynykBPS, HartmannKleihausKunz} or dilaton
fields~\cite{LM93dilaton, GMO93elud}.

In fact, black-hole solutions with hair were already known before
1989: in 1982, Gibbons found a black-hole solution with a non-trivial
dilaton field, within a model occurring in the low energy limit of
$N=4$ supergravity~\cite{GWG82dil}.

While the above counterexamples to the no-hair conjecture consist of
static, spherically-symmetric configurations, there exists numerical
evidence that static black holes are not necessarily spherically
symmetric~\cite{BKJK97BH1, Horowitz2011}; in fact, they might not even
need to be axisymmetric~\cite{RW95nonax}. Moreover, some studies also
indicate that non-rotating black holes need not be
static~\cite{BHSV97}. The rich spectrum of stationary--black-hole
configurations demonstrates that the matter fields are by far more
critical to the properties of black-hole solutions than expected for a
long time. In fact, the proof of the uniqueness theorem is, at least
in the axisymmetric case, heavily based on the fact that the
Einstein--Maxwell equations in the presence of a Killing symmetry form
a $\sigma$-model, effectively coupled to three-dimensional
gravity~\cite{NK69sigma}. ($\sigma$-models are a special case of
\emph{harmonic maps}, and we will use both terminologies
interchangeably in our context.) Since this property is not shared by
models with non-Abelian gauge fields~\cite{BH97PRD}, it is, with
hindsight, not too surprising that the Einstein--Yang--Mills system
admits black holes with hair.

However, there exist other black hole solutions, which are likely
to be subject to a generalized version of the uniqueness theorem.
These solutions appear in theories with self-gravitating massless
scalar fields (moduli) coupled to Abelian gauge fields. The
expectation that uniqueness results apply to a variety of these
models arises from the observation that their dimensional reduction
(with respect to a Killing symmetry) yields a $\sigma$-model with
symmetric target space (see, e.g.~\cite{BMG88KK, CG96, GL97},
and references therein).

\subsection{Organization}
\label{subsec-INT-ORG}

The purpose of this text is to review some features of
\emph{four-dimensional stationary asymptotically-flat} black-hole
spacetimes. Some black-hole solutions with non-zero cosmological
constant can be found in~\cite{VSLHB96, BHLSV96, WinstanleyEYM,
  SarbachWinstanely, RaduWinstaneley, BaxterWinstanley}. It should be
noted that the discovery of five-dimensional black rings by Emparan
and Reall~\cite{EmparanReall} has given new life to the overall
subject (see~\cite{EmparanReallReview, EmparanReallLR} and references
therein) but here we concentrate on four-dimensional spacetimes with
mostly classical matter fields.

For detailed introductions into the subject we refer to
Chandrasekhar's book on the mathematical theory of black
holes~\cite{SC91BK}, the classic textbook by Hawking and
Ellis~\cite{HE73LSS}, Carter's review~\cite{BC87CAR}, Chapter~12 of
Wald's book~\cite{BW84BK}, the overview~\cite{Chrusciel:2002mi} and
the monograph~\cite{MH96LN}.

The first part of this report is intended to provide a guide to the
literature, and to present some of the main issues, without going into
technical details. We start by collecting the significant definitions
in Section~\ref{SPrelim}. We continue, in Section~\ref{sect-CLASS}, by
recalling the main steps leading to the uniqueness theorem for
electro-vacuum black-hole spacetimes. The classification scheme
obtained in this way is then reexamined in the light of solutions,
which are \textit{not} covered by no-hair theorems, such as stationary
Kaluza--Klein black holes (Section~\ref{sec-KK}) and the
Einstein--Yang--Mills black holes (Section~\ref{sect-BEM}).

The second part reviews the main structural properties of stationary
black-hole spacetimes. In particular, we discuss the dimensional
reduction of the field equations in the presence of a Killing symmetry
in more detail (Section~\ref{sec-SST}). For a variety of matter
models, such as self-gravitating Abelian gauge fields, the reduction
yields a $\sigma$-model, with symmetric target manifold, coupled to
three-dimensional gravity. In Section~\ref{sect-ACS} we discuss some
aspects of this structure, namely the Mazur identity and the
quadratic mass formulae, and we present the Israel--Wilson class of
metrics.

The third part is devoted to stationary and axisymmetric black-hole
spacetimes (Section~\ref{sect-SAST}). We start by recalling the
circularity problem for non-Abelian gauge fields and for scalar
mappings. The dimensional reduction with respect to the second Killing
field leads to a boundary value problem on a fixed, two-dimensional
background. As an application, we outline the uniqueness proof for the
Kerr--Newman metric.


\newpage

\section{Definitions}
 \label{SPrelim}

It is convenient to start with definitions, which will be grouped
together in separate sections.

\subsection{Asymptotic flatness}
 \label{Saf}

We will mostly be concerned with asymptotically-flat black holes. A
spacetime $(\mcM,\fourgx)$ will be said to possess an
\emph{asymptotically-flat end \/}%
\index{asymptotic flatness}%
if  $\mcM$ contains a spacelike
hypersurface $\Mtext$ diffeomorphic to $\R^n\setminus B(R)$, where
$B(R)$ is an open coordinate ball of radius $R$, with the following
properties: there exists a constant $\alpha>0$ such that, in local
coordinates on $\Mtext$ obtained from $\R^n\setminus B(R)$, the metric
$\threeg$ induced by $\fourgx$ on $\Mtext$, the extrinsic curvature
tensor $K_{ij}$ of $\Mtext$, and the electromagnetic potential
$A_{\mu}$ satisfy the fall-off conditions
\begin{equation}
\label{falloff1}
  \threeg_{ij}-\delta_{ij}=O_{\diffk}(r^{-\alpha}) \,, \qquad
  K_{ij}=O_{{\diffk}-1}(r^{-1-\alpha}) \,,
\end{equation}
and
\begin{equation}
\label{Adecay}
  A_{\mu}=O_{\diffk}(r^{-{\alpha}}) \,,
\end{equation}
for some ${\diffk} > 1$, where we write $f=O_{\diffk}(r^{\alpha})$ if
$f$ satisfies
\begin{equation}
\label{okdef}
  \partial_{{i}_1}\ldots\partial_{{i}_\ell}
f=O(r^{\alpha-\ell})\,, \quad 0\le \ell \le {\diffk} \,.
\end{equation}

\subsection{Kaluza--Klein asymptotic flatness}
\label{subsec-KKdefxx}

There exists a generalization of the notion of asymptotic flatness,
which is relevant to both four- and higher-dimensional gravitation. We
shall say that $\hyp_{ext}$ is a \emph{Kaluza--Klein asymptotic end}
if $\hyp_{\ext}$ is diffeomorphic to $\left(\R^{\Ldim}\setminus\bar
B(R)\right)\times {\Compact}$, where $\bar B(R)$ is a closed
coordinate ball of radius $R$ and ${\Compact}$ is a compact manifold
of dimension $s\geq 0$; a spacetime containing such an end is said to
have ${\Ldim}+1$ asymptotically-large dimensions. Let $\mathring h$
be a fixed Riemaniann metric on ${\Compact}$, and let $\mathring
g=\delta\oplus\mathring h$, where $\delta$ is the Euclidean metric on
$\R^{\Ldim}$. A spacetime $(\mcM,\fourgx)$ containing such an end
will be said to be \emph{Kaluza--Klein asymptotically flat}, or
\emph{$KK$-asymptotically flat} if, for some $\alpha>0$, the metric
$\threeg$ induced by $\fourgx$ on $\Mtext$ and the extrinsic curvature
tensor $K_{ij}$ of $\Mtext$, satisfy the fall-off conditions
\begin{equation}
\label{falloff1KK}
  \threeg_{ij}-\mathring g_{ij}=O_{\diffk}(r^{-\alpha-l}) \,, \qquad
  K_{ij}=O_{{\diffk}-1}(r^{-1-\alpha-l}) \,,
\end{equation}
where, in this context, $r$ is the radius in $\R^{\Ldim}$\ and we
write $f=O_{\diffk}(r^{\alpha})$ if $f$ satisfies
\begin{equation}
\label{KKdecay}
  \mathring D_{i_1}\ldots \mathring D_{i_l} f=O(r^{\alpha-l})\,, \quad 0\le \ell \le {\diffk}\,,
\end{equation}
with $\mathring D$ the Levi-Civita connection of $\mathring g$.

\subsection{Stationary metrics}

An asymptotically-flat, or $KK$-asymptotically-flat, spacetime
$(\mcM,\fourgx)$ will be called \emph{stationary} if there exists on
$\mcM$ a complete Killing vector field $\changedX $, which is
\emph{timelike} in the asymptotic region $\Sext$; such a Killing
vector will be sometimes called \emph{stationary} as well. In fact,
in most of the literature it is implicitly assumed that stationary
Killing vectors satisfy $\fourgx(\changedX ,\changedX)<-\epsilon<0$
for some $\epsilon$ and for all $r$ large enough. This uniformity
condition excludes the possibility of a timelike vector, which
asymptotes to a null one. This involves no loss of generality in
well-behaved asymptotically-flat spacetimes: indeed, this uniform
timelikeness condition always holds for Killing vectors, which are
timelike for all large distances if the conditions of the positive
energy theorem are met~\cite{ChBeig1, ChMaerten}.

In electrovacuum, as part of the definition of stationarity it is also
required that the Maxwell field be invariant with respect to
$\changedX$, that is
\begin{equation}
\label{FXinv}
L_{\changedX} F\equiv0 \,.
\end{equation}

Note that this definition assumes that the Killing vector $\changedX$
is \emph{complete}, which means that for every $p\in \mcM$ the orbit
$\phi_t[\changedX ](p)$ of $\changedX $ is defined for all $t\in
\R$. The question of completeness of Killing vectors is an important
issue, which needs justifying in some steps of the uniqueness
arguments~\cite{ChCompleteness, PC96HPA}.

In regions where $\changedX$ is timelike, there exist local
coordinates in which the metric takes the form
\begin{equation}
\label{gme1}
 \fourgx =
 -V^2(dt+\underbrace{\theta_i dx^i}_{=:\theta})^2 +
 \underbrace{\threeg_{ij}dx^i dx^j}_{=:\threeg} \,,
\end{equation}
with
\begin{equation}
\label{foff0}
 \changedX =\partial_t \quad \Longrightarrow \quad
     \partial_t V = \partial_t \theta_i = \partial_t \threeg_{ij}=0 \,.
\end{equation}
Such coordinates exist globally on asymptotically-flat ends, and if
the Einstein--Maxwell equations hold, one can also obtain
there~\cite[Section 1.3]{PC94DG}, in dimension 3+1,
\begin{equation}
\label{foff}
 \threeg_{ij}-\delta_{ij}=O_{\infty}(r^{-1})\,, \quad
  \theta_{i} =O_{\infty}(r^{-1})\,, \quad V-1= O_{\infty}(r^{-1}) \,,
\end{equation}
and
\begin{equation}
\label{foff A}
 \quad A_{\mu}=O_{\infty}(r^{-1}) \,,
\end{equation}
where the infinity symbol means that~\eqref{okdef} holds for
arbitrary ${\diffk}$.

\subsection{Domains of outer communications, event horizons}
\label{sSdoc}

For $t\in {\R}$ let $\phi_t[\changedX ]:\mcM\to \mcM$ denote the
one-parameter group of diffeomorphisms generated by $\changedX $; we
will write $\phi_t$ for $\phi_t[\changedX ]$
whenever ambiguities are unlikely to occur.

Recall that $I^-(\Omega)$, respectively $J^-(\Omega)$, is the set
covered by past-directed timelike, respectively causal, curves
originating from $\Omega$, while $\dot I^- $ denotes the boundary of
$I^-$, etc. The sets $I^+$, etc., are defined as $I^-$, etc., after
changing time-orientation. See~\cite{HE73LSS, BeemEhrlichEasley,
  BONeill, MinguzziSanchez, PenroseDiffTopo, ChCausality} and
references therein for details of causality theory.

\begin{figure}[t]
\begin{center} { \psfrag{Mext}{\Large$\,\Mext$}
 \psfrag{H}{ } \psfrag{B}{ }
 \psfrag{pasthorizon}{\Large $\!\!\!\!\!{\partial I^+(\Mext)}$ }
 \psfrag{pSigma}{$\!\!\pohyp\qquad\phantom{xxxxxx}$}
 \psfrag{Sigma}{\Large $\!\Sext$}
  \psfrag{toto}{\Large$\!\!\!\!\!\!\!\!\!\!\!\!\!\!\!\!\! I^-(\Mext)$}
  \psfrag{S}{}
  \psfrag{future}{\Large $\!\!\!\!\!{I^+(\Mext)}$}
 \psfrag{H'}{ } \psfrag{W}{$\mathcal{W}$}
 \psfrag{scriplus} {} 
 \psfrag{scriminus} {} 
  \psfrag{i0}{}
 \psfrag{i-}{ } \psfrag{i+}{}
  \psfrag{E+}{\Large $\!\!\!\!\!{\partial I^-(\Mext)}$}
 \resizebox{2.3in}{!}{\includegraphics{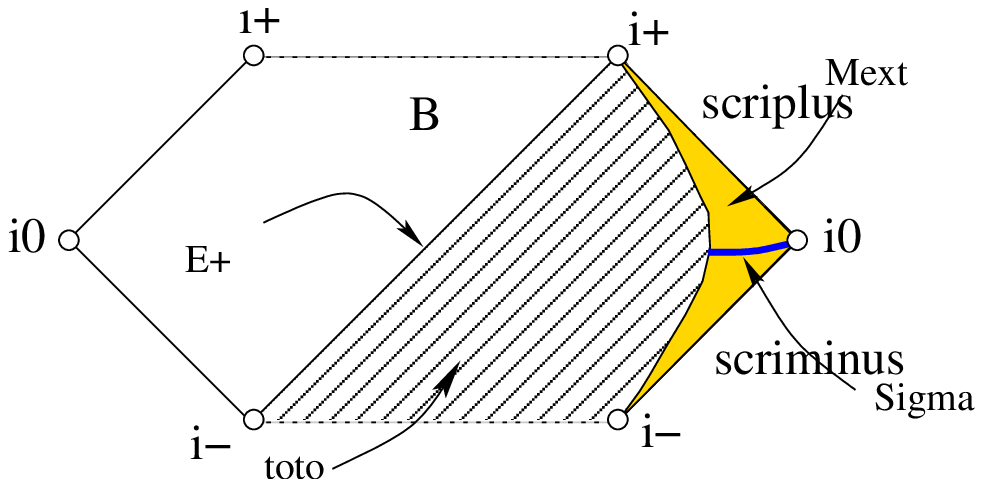}}
 \resizebox{2.3in}{!}{\includegraphics{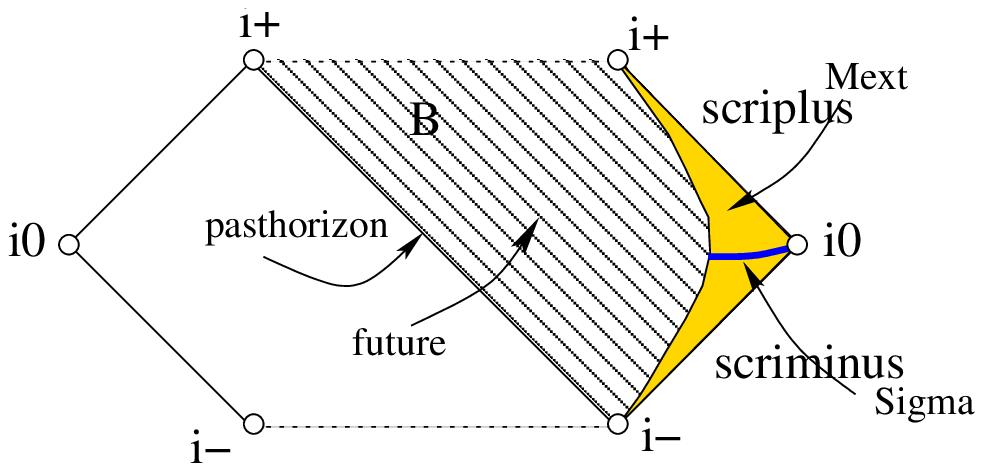}}
 }
 \caption{$\Sext$, $\Mext$, together with the future and the past of $\Mext$. One has $\Mext\subset I^\pm(\Mext)$, even
 though this is not immediately apparent from the figure.
 The domain of outer communications is the intersection $ I^+(\Mext)\cap I^-(\Mext)$, compare Figure~\ref{fregu}.
 \label{FPast}}
 \end{center}
\end{figure}

Consider an asymptotically-flat, or $KK$-asymptotically-flat, spacetime
with a Killing vector $\changedX$, which is timelike on the asymptotic
end $\Sext$.
\index{$I^{\pm}$, $\dot I^\pm$}%
\index{$J^{\pm}$, $\dot J^\pm$}%
The exterior region $\Mext$
\index{$\Mext$}
and the \emph{domain of outer communications}%
\index{domain of outer communications}
$\doc$, for which we will also use the abbreviation d.o.c.,  are then defined as (see Figure~\ref{FPast})

\begin{equation}
\label{docdef}
 \doc = I^+(\underbrace{\cup_t \phi_t(\Sext )}_{=:\Mext})\cap
 I^-(\cup_t \phi_t (\Sext )) \,.
\end{equation}
\index{$\doc$}%
The \emph{black-hole region} $\mcB$%
\index{black hole!region}
and the \emph{black-hole event horizon} $\mcH^+$
are defined as
$$
 \mcB= \mcM\setminus I^-(\Mext)\,,\quad \mcH^+=\partial \mcB \,.
$$
The \emph{white-hole region} $\mcW$ and the \emph{white-hole event horizon} $\mcH^-$%
\index{white hole event horizon}
are defined as above after changing time orientation:
$$
 \mcW= \mcM\setminus I^+(\Mext)\,,\quad \mcH^-=\partial \mcW
 \,, \quad \mcH=\mcH^+\cup \mcH^-\,.
$$
\index{$\mcH^\pm$}
It follows that the boundaries of $\doc $ are included in the event
horizons. We set
\begin{equation}
\label{epm}
  \mcE^\pm = \partial \doc \cap I^\pm (\Mext)
 \,, \qquad \mcE=\mcE^+\cup \mcE^- \,.
\end{equation}
\index{$\mcE^\pm$}%

There is considerable freedom in choosing the asymptotic region
$\Sext$. However, it is not too difficult to show that $I^\pm
(\Mext)$, and hence $\doc$, $\mcH^\pm$ and $\mcE^\pm$, are independent
of the choice of $\Sext$ whenever the associated $\Mext$'s overlap.

By standard causality theory, an event horizon is the union of
Lipschitz null hypersurfaces. It turns out that event horizons in
stationary spacetimes satisfying energy conditions are as smooth as
the metric allows~\cite{ChCo, ChDGH}; thus, smooth if the metric is
smooth, analytic if the metric is.

\subsection{Killing horizons}
\label{ssKH}

A null embedded
hypersurface, invariant under the flow of a Killing vector
$\changedX $, which coincides with a connected component of the set
$$
\mcrealN[\changedX ]:= \{\fourgx(\changedX ,\changedX )=0\,,\ \changedX \ne 0\}
 \,,
$$
is called a \emph{Killing horizon} associated to $\changedX $. We will
often write $\mcN[\changedX]$ for $\mcrealN[\changedX]$, whenever
$\mcrealN[\changedX]$ is a Killing horizon.

\subsubsection{Bifurcate Killing horizons}
\label{ssBKH}

The Schwarzschild black hole has an event horizon with a specific
structure, which is captured by the following definition: A set is
called a \emph{bifurcate Killing horizon} if it is the union of a a
smooth spacelike submanifold $S$ of co-dimension two, called the
\emph{bifurcation surface}, on which a Killing vector field $k$
vanishes, and of four smooth null embedded hypersurfaces obtained by
following null geodesics in the four distinct null directions normal
to $S$.

For example, the Killing vector $x\partial_t+t\partial_x$ in Minkowski
spacetime has a bifurcate Killing horizon, with the bifurcation
surface $\{t=x=0\}$. As already mentioned, another example is given by
the set $\{r=2m\}$ in Schwarzschild--Kruskal--Szekeres spacetime with
positive mass parameter $m$.

In the spirit of the previous definition, we will refer to the union
of two null hypersurfaces, which intersect transversally on a
2-dimensional spacelike surface as a \emph{bifurcate null surface}.

The reader is warned that a bifurcate Killing horizon is \emph{not} a
Killing horizon, as defined in Section~\ref{ssKH}, 
since the
Killing vector vanishes on $S$. If one thinks of $S$ as \emph{not}
being part of the bifurcate Killing horizon, then the resulting set is
again\emph{ not} a Killing horizon, since it has more than one
component.

\subsubsection{Killing prehorizons}
 \label{ss12VI.1}

One of the key steps of the uniqueness theory, as described in
Section~\ref{sect-CLASS}, forces one to consider ``horizon
candidates'' with local properties similar to those of a proper event
horizon, but with global behavior possibly worse: A connected, not
necessarily embedded, null hypersurface
$\mcN_0\subset\mcrealN[\changedX ]$ {to which $\changedX $ is tangent}
is called a \emph{Killing prehorizon}. In this terminology, a {Killing
  horizon} is a \emph{Killing prehorizon}, which forms a
\emph{embedded} hypersurface, which \emph{coincides} with a connected
component of $\mcrealN[\changedX ]$. The Minkowskian Killing vector
$\partial_t-\partial_x$ provides an example where $\mcrealN$ is not a
hypersurface, with every hyperplane $t+x=\const $ being a prehorizon.

The Killing vector $\changedX =\partial_t +Y$ on $\R\times \T^n$,
equipped with the flat metric, where $\T^n$ is an $n$-dimensional
torus, and where $Y$ is a unit Killing vector on $\T^n$ with dense
orbits, admits prehorizons, which are not embedded. This last example
is globally hyperbolic, which shows that causality conditions are not
sufficient to eliminate this kind of behavior.

Of crucial importance to the zeroth law of black-hole physics (to be
discussed shortly) is the fact that the $(k,k)$-component of the Ricci
tensor vanishes on horizons or prehorizons,
\begin{equation}
R(k,k) = 0 \quad \mbox{on} \quad H[k] \,.
\label{KH-4}
\end{equation}
This is a simple consequence of the Raychaudhuri equation.

The following two properties of Killing horizons and prehorizons play
a role in the theory of stationary black holes:

\begin{itemize}

\item A theorem due to Vishveshwara~\cite{CVV68} gives a
  characterization of the Killing horizon $H[k]$ in terms of the twist
  $\omega$ of $k$:\epubtkFootnote{See, e.g.,~\cite{DFCC90}, p.~239
    or~\cite{MH96LN}, p.~92 for the proof.} A connected component of
  the set $\mcrealN:=\{\sprod{k}{k} = 0\,,\ k\ne 0\}$ is a
  (non-degenerate) Killing horizon whenever
  \begin{equation}
    \omega = 0 \quad \mbox{and} \quad
    i_{k} \D k \neq 0 \quad \mbox{on} \quad \mcrealN\,.
    \label{KH-2}
  \end{equation}

\item The following characterization of Killing prehorizons is
  often referred to as the Vishveshwara--Carter  
  Lemma~\cite{BC73BH, BC69KH} (compare \cite[Addendum]{Chstatic}): Let
  $(M,\fourg)$ be a smooth spacetime with complete, static Killing
  vector $k$.  Then the set $\{p\in \mcM \, |\, \sprod{k}{k}|_p =
  0\,,\ k(p)\ne 0\} $ is the union of integral leaves of the
  distribution $k^{\perp}$, which are totally geodesic within
  $M\setminus \{k = 0\}$.

\end{itemize}

\subsubsection{Surface gravity: degenerate, non-degenerate and mean-non-degenerate horizons}
\label{ssSG}

An immediate consequence of the definition of a Killing horizon or
prehorizon is the proportionality of $k$ and $\D N$ on $H[k]$, where
%
$$
 N:=\fourg(k,k) \,.
$$
This follows, e.g., from $\sprod{k}{\D N} = 0$, since $L_{k}N=0$, and
from the fact that two orthogonal null vectors are proportional. The
observation motivates the definition of the \textit{surface gravity}
$\kappa$ of a Killing horizon or prehorizon $\mcN[\changedX]$, through
the formula
\begin{equation}
  \label{kdef0}
  d\left(\fourgx(\changedX ,\changedX )\right)|_\mcN = -2\kappa \changedX \ ,
\end{equation}
where we use the same symbol $k$ for the covector $
\fourgx_{\mu\nu}\,\changedX ^\nu dx^\mu$ appearing in the right-hand
side as for the vector $k^\mu \partial_\mu$.

The Killing equation implies $\D N = -2 \nabla_{k} k $; we see that
the surface gravity measures the extent to which the parametrization
of the geodesic congruence generated by $k$ is not affine.

A fundamental property is that {the surface gravity} $\kappa$ is
constant over horizons or prehorizons in several situations of
interest. This leads to the intriguing fact that the surface gravity
plays a similar role in the theory of stationary black holes as the
temperature does in ordinary thermodynamics. Since the latter is
constant for a body in thermal equilibrium, the result
\begin{equation}
\kappa = \mbox{constant \quad on} \quad H[k]
\label{KH-5}
\end{equation}
is usually called the zeroth law of black-hole
physics~\cite{BCH73laws}.

The constancy of $\kappa$ holds in vacuum, or for matter fields
satisfying the dominant-energy condition, see, e.g.,
\cite[Theorem~7.1]{MH96LN}. The original proof of the zeroth
law~\cite{BCH73laws} proceeds as follows: First, Einstein's equations
and the fact that $R(k,k)$ vanishes on the horizon imply that $T(k,k)
= 0$ on $H[k]$. Hence, the vector field $T(k):=T^{\mu}{}_{ \nu} k^{\nu
}\partial_{x^\mu}$ is perpendicular to $k$ and, therefore, space-like
(possibly zero) or null on $H[k]$. On the other hand, the dominant
energy condition requires that $T(k)$ is zero, time-like or
null. Thus, $T(k)$ vanishes or is null on the horizon. Since two
orthogonal null vectors are proportional, one has, using Einstein's
equations again, $k \wedge R(k)= 0$ on $H[k]$, where $R(k)=
R_{\mu\nu}k^\mu dx^\nu$. The result that $\kappa$ is constant over
each horizon follows now from the general property (see, e.g.,
\cite{BW84BK})
\begin{equation}
k \wedge \D \kappa = - k \wedge R(k) \quad \mbox{on} \quad
H[k] \, .
\label{KH-6}
\end{equation}

The proof of~\eq{KH-5} given in~\cite{BW84BK} generalizes to all
spacetime dimensions $n+1\ge 4$; the result also follows in all
dimensions from the analysis in~\cite{HIW} when the horizon has
compact spacelike sections.

By virtue of Eq.~(\ref{KH-6}) and the identity $\D \omega = \ast [k
  \wedge R(k)]$, the zeroth law follows if one can show that the twist
one-form is closed on the horizon~\cite{IRRW95}:
\begin{equation}
\left[\D \omega \right]_{H[k]} = 0 \; \Longrightarrow \;
\kappa = \mbox{constant \quad on} \quad H[k].
\label{KH-7}
\end{equation}
While the original proof of the zeroth law takes advantage of
Einstein's equations and the dominant energy condition to conclude
that the twist is closed, one may also achieve this by requiring that
$\omega$ vanishes identically, which then proves the zeroth law under
the second set of hypotheses listed below.  This is obvious for static
configurations, since then $k$ has vanishing twist by definition.

Yet another situation of interest is a spacetime with two commuting
Killing vector fields $k$ and $m$, with a Killing horizon $H[\xi]$
associated to a Killing vector $\xi = k + \Omega m$. Such a spacetime
is said to be \emph{circular} if the distribution of planes spanned by
$k$ and $m$ is hypersurface-orthogonal. Equivalently, the metric can
be locally written in a 2+2 block-diagonal form, with one of the
blocks defined by the orbits of $k$ and $m$. In the circular case one
shows that $\sprod{m}{\omega_\xi} = \sprod{\xi}{\omega_m} = 0$ implies
$\D \omega_{\xi} = 0$ on the horizon generated by $\xi$;
see~\cite{MH96LN}, Chapter~7 for details.

A significant observation is that of Kay and Wald~\cite{KayWald}, that
$\kappa$ must be constant on bifurcate Killing horizons, regardless of
the matter content. This is proven by showing that the derivative of
the surface gravity in directions tangent to the bifurcation surface
vanishes. Hence, $\kappa$ cannot vary between the null-generators. But
it is clear that $\kappa$ is constant along the generators.

Summarizing, each of the following hypotheses is sufficient to prove
that $\kappa$ is constant over a Killing horizon defined by $k$:

\begin{enumerate}[(i)]
\item The dominant energy condition holds;
\item the domain of outer communications is static;
\item the domain of outer communications is circular;
\item $H[k]$ is a \textit{bifurcate} Killing horizon.
\end{enumerate}

See~\cite{IRRW95} for some further observations concerning \eq{KH-5}.

A Killing horizon is called \emph{degenerate} if $\kappa$ vanishes,
and \emph{non-degenerate} otherwise.

As an example, in Minkowski spacetime, consider the Killing vector
$\KV=x\partial_t+ t\partial_x$. We have
$$
d(\fourgx(\KV,\KV))=d(-x^2+t^2)=2 (-xdx+tdt) \,,
$$
which equals twice $\KV^\flat:= \fourgx_{\mu\nu}\KV^\mu dx^\nu$ on
each of the four Killing horizons
$$
\mcN(\KV)_{\epsilon \delta} :=\{
t=\epsilon x\,, \delta t >0\}\,, \quad \epsilon, \delta \in \{\pm 1\} \,.
$$
On
the other hand, for the Killing vector
\begin{equation}
\label{nullboost}
 \changedX = y\partial_t + t\partial_y + x \partial_y - y \partial_x =
 y\partial_t + (t+x)\partial_y - y \partial_x
\end{equation}
one obtains
$$
d(\fourgx(\changedX,\changedX))=2(t+x) (dt+dx)
 \,,
$$
which vanishes on each of the Killing horizons $\{t=-x\,, y\ne 0\}$.
This shows that the same null surface can have zero or non-zero
values of surface gravity, depending upon which Killing vector has
been chosen to calculate $\kappa$.

A key theorem of R\'acz and Wald~\cite{IRRW95} asserts that
non-degenerate horizons (with a compact cross section and constant
surface gravity) are ``essentially bifurcate'', in the following
sense: Given a spacetime with such a non-degenerate Killing horizon,
one can find another spacetime, which is locally isometric to the
original one in a one-sided neighborhood of a subset of the horizon,
and which contains a bifurcate Killing horizon. The result can be made
global under suitable conditions.

The notion of \emph{average} surface gravity can be defined for null
hypersurfaces, which are not necessarily Killing horizons:
Following~\cite{VinceJimcompactCauchyCMP}, near a smooth null
hypersurface $\mcrealN$ one can introduce \emph{Gaussian null
  coordinates}, in which the metric takes the form
\begin{equation}
\label{GNC1} \fourgx=r \varphi dv^2 + 2dv dr + 2r h_a dx^a dv + h_{ab}dx^a
dx^b\,.
\end{equation}
The null hypersurface $\mcrealN$ is given by the equation $\{r=0\}$;
when it corresponds to an event horizon, by replacing $r$ by $-r$ if
necessary we can, without loss of generality, assume that $r>0$ in the
domain of outer communications. Assuming that $\mcrealN$ admits a
smooth compact cross-section $S$, the \emph{average surface gravity}
$\langle \kappa\rangle_S$
is defined as
\begin{equation}
\label{asg}
 \langle \kappa\rangle_S=-\frac 1 {|S|}\int_S \varphi d\mu_h
 \,,
\end{equation}
where $d\mu_h$ is the measure induced by the metric $h$ on $S$, and
$|S|$ is the area of $S$. We emphasize that this is defined regardless
of whether or not the hypersurface is a Killing horizon; but if it is
with respect to a vector $\changedX$, and if the surface gravity
$\kappa$ of $\changedX$ is constant on $S$, then $\langle
\kappa\rangle_S$ equals $\kappa$.

A smooth null hypersurface, not necessarily a Killing horizon, with a
smooth compact cross-section $S$ such that $\langle
\kappa\rangle_S\neq 0$ is said to be \emph{mean non-degenerate}.

Using general identities for Killing fields (see, e.g.,~\cite{MH96LN},
Chapter~2) one can derive the following explicit expressions for
$\kappa$:
\begin{equation}
\kappa^2 = - \lim_{N\to 0}
\left[\frac{1}{N}\sprod{\nabla_{k} k}{\nabla_{k}
k}\right] = -
\left[\frac{1}{4} \Delta_{\fourg} N \right]_{H[k]} \,,
\label{KH-3}
\end{equation}
where $\Delta_{\fourg} $ denotes the Laplace--Beltrami operator of the
metric $\fourg$. Introducing the four velocity $u = k/\sqrt{-N}$ for a
time-like $k$, the first expression shows that the surface gravity is
the limiting value of the force applied at infinity to keep a unit
mass at $H[k]$ in place: $\kappa = \lim (\sqrt{-N}|a|)$, where $a =
\nabla_{u}u$ (see, e.g.,~\cite{BW84BK}).

\subsection{\textit{I}\super{+}-regularity}
\label{S3VI11.5}

\begin{figure}[ht]
 \begin{center} { \psfrag{Mext}{$\phantom{x,}\Mext$}
 \psfrag{H}{ } \psfrag{B}{ }
 \psfrag{H}{ }
 \psfrag{pSigma}{$\!\!\pohyp\qquad\phantom{xxxxxx}$}
 \psfrag{Sigma}{ $\hyp$ }
 \psfrag{toto}{$\!\!\!\!\!\!\!\!\!\!\doc$}
 \psfrag{S}{}
 \psfrag{H'}{ } \psfrag{W}{$\mathcal{W}$}
 \psfrag{scriplus} {} 
 \psfrag{scriminus} {} 
 \psfrag{i0}{}
 \psfrag{i-}{ } \psfrag{i+}{}
 \psfrag{E+}{ $\phantom{.}{\mycal E}^+$}
 \resizebox{3in}{!}
 {\includegraphics{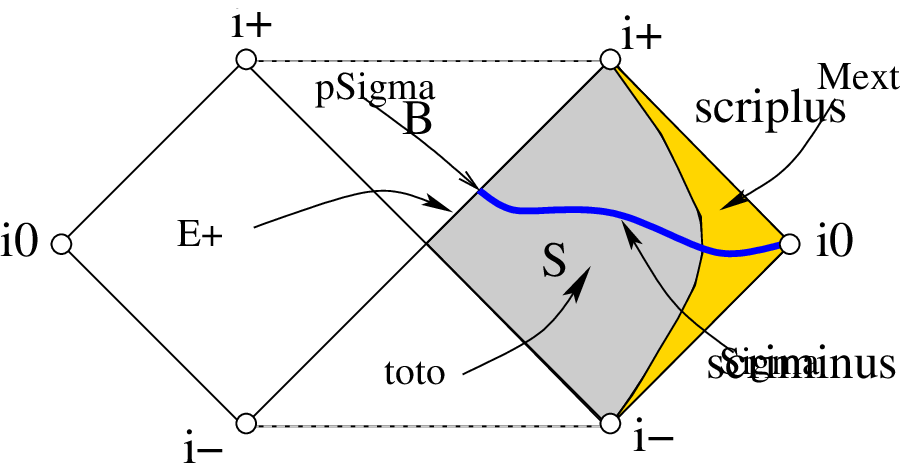}}
 }
 \caption{The hypersurface $\hyp$ from the definition of \regular ity.
 \protect\label{fregu}}
 \end{center}
 \end{figure}

The classification theory of stationary black holes requires that the
spacetime under consideration satisfies various global regularity
conditions. These are captured by the following definition:

\begin{Definition}
\label{Dmain}
Let $(\mcM,\fourgx)$ be a spacetime containing an asymptotically-flat
end, or a $KK$-asymp\-toti\-cally-flat end $\Sext$, and let $\changedX $
be a stationary Killing vector field on $\mcM$. We will say that
$(\mcM,\fourgx,\changedX)$ is $I^+$-regular%
\index{$I^+$-regular}
if ${\changedX}$ is complete, if the domain of outer communications
$\doc$ is globally hyperbolic, and if $\doc$ contains a spacelike,
connected, acausal hypersurface $\hyp\supset\Sext $,%
\  the closure $\ohyp $ of which is a topological manifold with boundary,
consisting of the union of a compact set and of a finite number of
asymptotic ends, such that the boundary $ \pohyp:= \ohyp \setminus \hyp$
is a topological manifold satisfying
\begin{equation}
\label{subs}
\pohyp \subset \mcE^+:= \partial \doc \cap I^+(\Mext) \,,
\end{equation}
with $\pohyp$ meeting every generator of $\mcE^+$ precisely once. (See
Figure~\ref{fregu}.)
\end{Definition}

The ``$I^+$'' of the name is due to the $I^+$ appearing in \eq{subs}.

Some comments about the definition are in order. First, one
requires completeness of the orbits of the stationary Killing
vector to have an action of $\R$ on $\mcM$ by
isometries. Next, global hyperbolicity of the domain
of outer communications is used to guarantee its simple connectedness,
to make sure that the area theorem holds, and to avoid
causality violations as well as certain kinds of naked
singularities in $\doc$. Further, the existence of a
well-behaved spacelike hypersurface is a prerequisite to any
elliptic PDEs analysis, as is extensively needed for the
problem at hand. The existence of compact cross-sections of the
future event horizon prevents singularities on the future part
of the boundary of the domain of outer communications, and
eventually guarantees the smoothness of that boundary. The requirement
Eq.~\eq{subs} might appear somewhat unnatural, as there are perfectly
well-behaved hypersurfaces in, e.g., the Schwarzschild spacetime,
which do not satisfy this condition, but there arise various technical
difficulties without this condition.
Needless to say, all those conditions are satisfied
by the Kerr--Newman and the Majumdar--Papapetrou (MP) solutions.

\newpage

\section{Towards a classification of stationary electrovacuum black hole
  spacetimes}
\label{sect-CLASS}
 
While the uniqueness theory for black-hole solutions of Einstein's
vacuum equations and the Einstein--Maxwell (EM) equations has seen
deep successes, the complete picture is nowhere settled at the time of
revising of this work. We know now that, under reasonable global
conditions (see Definition~\ref{Dmain}), the domains of dependence of
analytic, stationary, asymptotically-flat electrovacuum black-hole
spacetimes with a connected non-degenerate horizon belong to the
Kerr--Newman family. The purpose of this section is to review the
various steps involved in the classification of electrovacuum
spacetimes (see Figure~\ref{fig:classification}). In Section~\ref{sect-BEM} 
we shall then comment on the validity of the partial results in the
presence of non-linear matter fields.

For definiteness, from now on we assume that all spacetimes are
$I^+$-regular. We note that the slightly weaker global conditions
spelled-out in Theorem~\ref{Tubhelvstatic} suffice for the analysis of
static spacetimes, or for various intermediate steps of the
uniqueness theory, but those weaker conditions are not known to
suffice for the Uniqueness Theorem~\ref{Tubh}.

The main task of the uniqueness program is to show that the domains of
outer communications of sufficiently regular stationary electrovacuum
black-hole spacetimes are exhausted by the Kerr--Newman or the
MP spacetimes.

The starting point is the smoothness of the event horizon; this is
proven in~\cite[Theorem~4.11]{ChCo}, drawing heavily on the results
in~\cite{ChDGH}.

One proves, next, that connected components of the event horizon are
diffeomorphic to $\bbbr \times S^2$. This was established
in~\cite{CW94TOP}, taking advantage of the topological censorship
theorem of Friedman, Schleich and Witt~\cite{FSW93};
compare~\cite{SH72BH} for a previous partial result. (Related versions
of the topology theorem, applying to globally-hyperbolic, not-necessarily-stationary, spacetimes, have been established by Jacobson
and Venkataramani~\cite{JV95TOP}, and by Galloway~\cite{Gallo93top,
  Gallo95top, Gallo96top, GW97top}; the strongest-to-date version,
with very general asymptotic hypotheses, can be found in~\cite{CGS}.)

\subsection{Static solutions}
\label{S3VI11.1}

A stationary spacetime is called \emph{static} if the Killing vector
$\changedX$ is hypersurface-orthogonal: this means that the
distribution of the hyperplanes orthogonal to $\changedX$ is
integrable. Equivalently,
$$
\changedX\wedge d\changedX = 0 \,.
$$
Here and elsewhere, by a common abuse of notation, we also write
$\changedX$ for the one-form associated with $\changedX$.

The results concerning static black holes are stronger than the
general stationary case, and so this case deserves separate
discussion. In any case, the proof of uniqueness for stationary black
holes branches out at some point and one needs to consider separately
uniqueness for static configurations.

In pioneering work, Israel showed that both static
vacuum~\cite{WI67vac} and electrovacuum~\cite{WI68evac} black-hole
spacetimes satisfying a set of restrictive conditions are spherically
symmetric. Israel's ingenious method, based on differential identities
and Stokes' theorem, triggered a series of investigations devoted to
the static uniqueness problem (see, e.g.,~\cite{MzH73vac, MzH74evac,
  DCR74, DCR77, WS85UT}). A breakthrough was made by Bunting and
Masood-ul-Alam~\cite{BMuA87UT}, who showed how to use the positive
energy theorem%
\epubtkFootnote{This theorem was first proven by Schoen
  and Yau~\cite{SY79PET, SY81PET} and somewhat later, using spinor
  techniques, by Witten~\cite{EW81PET}
  (compare~\cite{TP82PET}). See~\cite{BartnikChrusciel2} for a version
  relevant to the uniqueness problem, which allows degenerate
  components of the event horizon.} to exclude non-connected
configurations (compare~\cite{Chstatic}).%
\epubtkFootnote{Non-existence of certain static $n$-body
  configurations (possibly, but not necessarily, black holes) was
  established in~\cite{Beig:2008qi, Beig:2009jd}). These results rely
  on the positive energy theorem and exclude, in particular, suitably
  regular configurations with a reflection symmetry across a
  noncompact surface, which is disjoint from the matter regions.}

The annoying hypothesis of analyticity, which was implicitly assumed
in the above treatments, has been removed in \cite{ChGstatic}. The
issue here is to show that the Killing vector field cannot become null
on the domain of outer communications. The first step to prove this is
the Vishveshwara--Carter lemma (see Section~\ref{ss12VI.1}
and~\cite{CVV68, BC69KH}), which shows that null orbits of static
Killing vectors form a \emph{prehorizon}, as defined in
Section~\ref{ss12VI.1}. To finish the proof one needs to show that
prehorizons cannot occur within the d.o.c.  This presents no
difficulty when analyticity is assumed. Now, analyticity of stationary
electrovacuum metrics is a standard property~\cite{MzH74evac, MzH}
when the Killing vector is timelike, but timelikeness throughout the
d.o.c.\ is not known yet at this stage of the argument. The
nonexistence of prehorizons within the d.o.c.\ for smooth metrics
requires more work, and is the main result in~\cite{ChGstatic}.

In the static \emph{vacuum} case the remainder of the argument can be
simplified by noting that there are no static solutions with
degenerate horizons, which have spherical
cross-sections~\cite{CRT}. This is not true anymore in the
electrovacuum case, where an intricate argument to handle
non-degenerate horizons is needed~\cite{CT} (compare~\cite{Ruback,
  Simon:elvac, MuA92UT, ChstaticelvacarxivErr} for previous partial
results).

All this can be summarized in the following classification theorem:

\begin{Theorem}
\label{Tubhelvstatic}
Let $(\mcM,\fourg)$ be an electrovacuum, four-dimensional spacetime
containing a spacelike, connected, acausal hypersurface $\hyp $, such
that $\ohyp $ is a topological manifold with boundary consisting of
the union of a compact set and of a finite number of asymptotically-flat ends. Suppose that there exists on $\mcM$ a complete
hypersurface-orthogonal Killing vector, that the domain of outer
communication $\doc$ is globally hyperbolic, and that $
\pohyp\subset\mcM\setminus \doc$. Then $\doc$ is isometric to the
domain of outer communications of a Reissner--Nordstr\"om or a
MP spacetime.
\end{Theorem}

\subsection{Stationary-axisymmetric solutions}
\label{S3VI11.1x}

\subsubsection{Topology}

A second class of spacetimes where reasonably satisfactory statements
can be made is provided by stationary-axisymmetric solutions. Here one
assumes from the outset that, in addition to the stationary Killing
vector, there exists a second Killing vector field. Assuming $I^+$-regularity, one can
invoke the positive energy theorem  to show~\cite{PC96ISO, ChBeig3} that some
linear combination of the Killing vectors, say
\newcommand{\myeta}{m}%
$\myeta$,  must have periodic orbits, and an \emph{axis of rotation},
i.e., a two-dimensional totally-geodesic submanifold of $\mcM$ on
which the Killing vector $\myeta$ vanishes. The description of the
quotient manifold is provided by the deep mathematical results
concerning actions of isometry groups of~\cite{Orlik, Raymond},
together with the simple-connectedness and structure
theorems~\cite{ChCo}. The bottom line is that the spacetime is the
product of $\R$ with $\R^3$ from which a finite number of aligned
balls, each corresponding to a black hole, has been removed. Moreover,
the $U(1)$ component of the group of isometries acts by rotations of
$\R^3$. Equivalently, the quotient space is a half-plane from which
one has removed a finite number of disjoint half-discs centered on
points lying on the boundary of the half-plane.

\subsubsection{Candidate metrics}
\label{ssCand}

The only known $I^+$-regular stationary and axisymmetric solutions of
the Einstein--Maxwell equations are the Kerr--Newman metrics and the
MP metrics. However, it should be kept in mind
that candidate solutions for non-connected black-hole configurations
exist:

First, there are the multi-soliton metrics constructed using inverse
scattering methods~\cite{BelinskiZakharov, BelinskiZakharov2}  (compare
\cite{Pomeransky}). Closely related (and possibly identical,
see~\cite{HerdeiroRebelo}), are the multi-Kerr solutions constructed by
successive B\"acklund transformations starting from Minkowski
spacetime; a special case is provided by the Neugebauer--Kramer
double-Kerr solutions~\cite{KramerNeugebauerDoubleKerr}. These are
explicit solutions, with the metric functions being rational functions
of coordinates and of many parameters. It is known that some subsets
of those parameters lead to metrics, which are smooth at the axis of
rotation, but one suspects that those metrics will be nakedly singular
away from the axis. We will return to that question in
Section~\ref{ss14VI11.3}.

Next, there are the solutions constructed by
Weinstein~\cite{Weinstein3}, obtained from an abstract existence
theorem for suitable harmonic maps. The resulting metrics are smooth
everywhere except perhaps at some components of the axis of
rotation. It is known that some Weinstein solutions have conical
singularities~\cite{Weinstein:trans, LiTian, HennigNeugebauer3, CENS}
on the axis, but the general case remains open.

Finally, the Israel--Wilson--Perj\'es (IWP) metrics~\cite{ZP71,
  IW72Class}, discussed in more detail in Section~\ref{subsec-IWC}, provide candidates for rotating generalizations of the
MP black holes. Those metrics are remarkable because
they admit nontrivial Killing spinors. The Killing vector obtained
from the Killing spinor is causal everywhere, so the horizons are
necessarily non-rotating and degenerate. It has been shown
in~\cite{CRTIWP} that the only regular IWP metrics
\emph{with a Killing vector timelike throughout the d.o.c.\ } are the
MP metrics. A strategy for a proof of timelikeness
has been given in~\cite{CRTIWP}, but the details have yet to be
provided. In any case, one expects that the only regular IWP metrics
are the MP ones.

Some more information concerning candidate solutions with
non-connected horizons can be found in Section~\ref{ss14VI11.3}.

\subsubsection{The reduction}
\label{sec-red}

Returning to the classification question, the analysis continues with
the \textit{circularity theorem} of Papapetrou~\cite{Papapetrou:ot} 
and Kundt and Tr\"umper~\cite{KT66CIRC} (compare~\cite{BC69KH}), which
asserts that, locally and away from null orbits, the metric of a
vacuum or electrovacuum spacetime can be written in a 2+2
block-diagonal form.

The next key observation of Carter is that the stationary and
axisymmetric EM equations can be reduced to a two-dimensional boundary
value problem~\cite{BC71bd} (see Sections~\ref{subsec-REHA}
and~\ref{subsec-BVCM} for more details), provided that the \emph{area
  density of the orbits of the isometry group can be used as a global
  spacelike coordinate} on the quotient manifold. (See
also~\cite{BC79GT} and~\cite{BC87CAR}.) Prehorizons intersecting the
d.o.c.\ provide one of the obstructions to this, and a heavy-duty
proof that such prehorizons do not arise was given in \cite{ChCo}; a
simpler argument has been provided in~\cite{ChGstatic}.

In essence, Carter's reduction proceeds through  a 
global manifestly--conformally-flat (``isothermal'') coordinate system
$(\rho,z)$ on the quotient manifold. One also needs to carefully
monitor the boundary conditions satisfied by the fields of
interest. The proof of existence of the $(\rho,z)$ coordinates, with
sufficient control of the boundary conditions so that the uniqueness
proof goes through, has been given in~\cite{ChCo}, drawing heavily
on~\cite{ChUone}, assuming that all horizons are non-degenerate. A
streamlined argument has been presented in~\cite{ChNguyen}, where the
analysis has also been extended to cover configurations with
degenerate components.

So, at this stage one has reduced the problem to the study of
solutions of harmonic-type equations on $\R^3 \setminus \mcA$, where
$\mcA$ is the rotation axis $\{x=y=0\}$, with precise boundary
conditions at the axis. Moreover, the solution is supposed to be
invariant under rotations. Equivalently, one has to study a set of
harmonic-type equations on a half-plane with specific singularity
structure on the boundary.

There exist today at least three arguments that finish the proof, to
be described in the following subsections.

\subsubsection{The Robinson--Mazur proof}
\label{ss11VI11.1}

In the vacuum case, Robinson was able to construct an amazing
identity, by virtue of which the uniqueness of the Kerr metric
followed~\cite{DCR75Ident}. The uniqueness problem with
electro-magnetic fields remained open until Mazur~\cite{PM82UT}
obtained a generalization of the Robinson identity in a systematic
way: The Mazur identity (see also~\cite{PM84a, PM84b, CarterCMP,
  BMG88KK, HY2, HY}) is based on the observation that the EM equations
in the presence of a Killing field describe a non-linear
$\sigma$-model with coset space $G/H = SU(1,2) / S(U(1) \times
U(2))$. The key to the success is Carter's idea to carry out the
dimensional reduction of the EM action with respect to the axial
Killing field. Within this approach, the Robinson identity loses its
enigmatic status -- it turns out to be the explicit form of the Mazur
identity for the vacuum case, $G/H = SU(1,1) / U(1)$.

Reduction of the EM action with respect to the time-like Killing field
yields, instead, $H = S(U(1,1) \times U(1))$, but the resulting
equations become singular on the ergosurface, where the Killing vector
becomes null.

More information on this subject is provided in
Sections~\ref{subsec-TMI} and \ref{subsec-DivId}.

\subsubsection{The Bunting--Weinstein harmonic-map argument}
\label{ss11VI11.2}

At about the same time, and independently of Mazur,
Bunting~\cite{GB83UT} gave a proof of uniqueness of the relevant
harmonic-map equations exploiting the fact that the target space for
the problem at hand is negatively curved. A further systematic PDE
study of the associated harmonic maps has been carried out by
Weinstein: as already mentioned, Weinstein provided existence results
for multi-horizon configurations, as well as uniqueness
results~\cite{Weinstein3}.

All the uniqueness results presented above require precise asymptotic
control of the harmonic map and its derivatives at the singular set
$\mcA$. This is an annoying technicality, as no detailed study of the
behavior of the derivatives has been presented in the literature. The
approach in~\cite[Appendix~C]{CLW} avoids this problem, by showing
that a pointwise control of the harmonic map is enough to reach the
desired conclusion.

For more information on this subject consult Section~\ref{ss14VI11.1}.

\subsubsection{The Varzugin--Neugebauer--Meinel argument}
\label{ss11VI11.3}

The third strategy to conclude the uniqueness proof has been advocated
by Varzugin~\cite{Varzugin1, Varzugin2} and, independently, by
Neugebauer and Meinel~\cite{Neugebauer:2003qe}. The idea is to exploit
the properties of the \emph{linear problem} associated with the
harmonic map equations, discovered by Belinski and
Zakharov~\cite{BelinskiZakharov, BelinskiZakharov2} (see
also~\cite{Pomeransky}). This proceeds by showing that a regular
black-hole solution must necessarily be one of the multi-soliton
solutions constructed by the inverse-scattering methods, providing an
argument for uniqueness of the Kerr solution within the class. Thus, one
obtains an explicit form of the candidate metric for solutions
with more components, as well as an argument for the non-existence of
two-component configurations~\cite{HennigNeugebauer3}
(compare~\cite{CENS}).

\subsubsection{The axisymmetric uniqueness theorem}

What has been said so far can be summarized as follows:

\begin{theorem}
\label{Tubhaxi}
Let $(\mcM,\fourg)$ be a stationary, axisymmetric asymptotically-flat,
$I^+$-regular, electrovacuum four-dimensional spacetime. Then the domain
of outer communications $\doc$ is isometric to one of the Weinstein
solutions. In particular, if the event horizon is connected, then
$\doc$ is isometric to the domain of outer communications of a
Kerr--Newman spacetime.
\end{theorem} 

\subsection{The no-hair theorem}
\label{s13VI11.1}

\subsubsection{The rigidity theorem}
\label{S3VI11.3}
 
Throughout this section we will assume that the spacetime is
$I^+$-regular, as made precise by Definition~\ref{Dmain}.

To prove uniqueness of connected, analytic, non-degenerate
configurations, it remains to show that every such black hole is either
static or axially symmetric. The first step for this is provided by
Hawking's \textit{strong rigidity theorem} (SRT)~\cite{HE73LSS,
  VinceJimcompactCauchyCMP, Ch:rigidity, FRW}, which relates the
global concept of the event horizon to the independently-defined, and
logically-distinct, local notion of the Killing horizon. Assuming
analyticity, SRT asserts that the event horizon of a
\textit{stationary} black-hole spacetime is a Killing horizon. (In
this terminology~\cite{MH96LN}, the \emph{weak rigidity theorem} is
the existence, already discussed above, of prehorizons for static or
stationary and axisymmetric configurations.)

A Killing horizon is called \emph{non-rotating} if it is generated by
the stationary Killing field, and \emph{rotating} otherwise. At this
stage the argument branches-off, according to whether at least one of
the horizons is rotating, or not.

In the rotating case, Hawking's theorem actually provides only a
second Killing vector field defined near the Killing horizon, and to
continue one needs to globalize the Killing vector field, to prove
that its orbits are complete, and to show that there exists a linear
combination of Killing vector fields with periodic orbits and an axis
of rotation. This is done in~\cite{Ch:rigidity}, assuming analyticity,
drawing heavily on the results in~\cite{Nomizu, ChCompleteness,
  PC96ISO}.

The existing attempts in the literature to construct a second Killing
vector field without assuming analyticity have only had limited
success. One knows now how to construct a second Killing vector in a
neighborhood of non-degenerate horizons for electrovacuum black
holes~\cite{AIK, IonescuKlainerman1, PinYuHawkingElVac}, but the
construction of a second Killing vector throughout the d.o.c.\ has
only been carried out for vacuum near-Kerr non-degenerate
configurations so far~\cite{AIK2} (compare~\cite{WongKerrNewman}).

In any case, sufficiently regular analytic stationary electro-vacuum
spacetimes containing a rotating component of the event horizon are
axially symmetric as well, regardless of degeneracy and connectedness
assumptions (for more on this subject see
Section~\ref{subsec-RigOpen}). One can then finish the uniqueness
proof using Theorem~\ref{Tubhaxi}. Note that the last theorem requires
neither analyticity nor connectedness, but leaves open the question of the
existence of naked singularities in non-connected candidate solutions.

In the non-rotating case, one continues by showing~\cite{CW94MAX} that
the domain of outer communications contains a maximal Cauchy
surface. This has been proven so far only for non-degenerate horizons,
and this is the only missing step to include situations with
degenerate components of the horizon. This allows one to prove the
\textit{staticity theorem}~\cite{SW92stat, SW93stat}, that the
stationary Killing field of a non-rotating, electrovacuum black-hole
spacetime is hypersurface orthogonal. (Compare~\cite{PH73ERGO,
  PH75BIF, HE73LSS, SH72BH} for previous partial results.) One can
then finish the argument using Theorem~\ref{Tubhelvstatic}.

\subsubsection{The uniqueness theorem}
 \label{subsect-CLASS-UT}

All this leads to the following precise statement, as proven
in~\cite{ChCo, ChNguyen} in vacuum and in~\cite{Costaelvac, ChNguyen}
in electrovacuum:

\begin{theorem}
\label{Tubh}
Let $(\mcM,\fourg)$ be a stationary, asymptotically-flat, $I^+$-regular,
electrovacuum, four-dimensional analytic spacetime. If the event
horizon is connected and \emph{either} mean non-degenerate \emph{or}
rotating, then $\doc$ is isometric to the domain of outer
communications of a Kerr--Newman spacetime.
\end{theorem}

The structure of the proof can be summarized in the flow chart of
Figure~\ref{fig:classification}. This is to be compared with
Figure~\ref{fig:classification2}, which presents in more detail the
weaker hypotheses needed for various parts of the argument. 

%
%

\epubtkImage{classification.png}{%
\begin{figure}[htbp]
\small
\begin{center}
\framebox[\textwidth]{Stationary, analytic, connected, mean non-degenerate, $I^+$-regular electrovacuum black-hole spacetime}\\
\vspace{.15cm}
[asymptotically time-like Killing field $k^{\mu}$]
\end{center}

\vspace{.35cm}
\begin{center}
\fbox{{\bf{Smoothness and topology theorems}} }\\
\vspace{.15cm}
\end{center}

\vspace{.35cm}
\begin{center}
\fbox{{\bf{STRONG RIGIDITY THM ({\boldmath$1^{\rm st}$} part)}} }
\end{center}

\begin{center}
\fbox{Event horizon $=$ Killing horizon $H[\xi]$}\\
\vspace{.15cm}
[null generator Killing field $\xi^{\mu}$]
\end{center}

\begin{tabbing}
\fbox{$H[\xi]$ non-rotating: $k^{\mu}k_{\mu}\!\mid_{H[\xi]}\equiv 0$}
\`\fbox{$H[\xi]$ rotating: $k^{\mu}k_{\mu}\!\mid_{H[\xi]} \not\equiv 0$}
\end{tabbing}

\vspace{.35cm}
\begin{tabbing}
\fbox{{\bf{STATICITY THM}}}
\`\fbox{{\bf{STRONG RIGIDITY THM ({\boldmath$2^{\rm nd}$} part)}}}
\end{tabbing}

\begin{tabbing}
\fbox{d.o.c.\ static }
\`\fbox{d.o.c.\ axisymmetric}
\end{tabbing}
\vspace{-.6cm}
\begin{tabbing}
[$k_{[\alpha}\nabla_{\beta}k_{\gamma]}=0$]
\`[$\exists$ Killing field $m^{\mu}$]
\end{tabbing}

\vspace{.35cm}
\begin{tabbing}
\fbox{{\bf{STATICITY THM ({\boldmath$2^{\rm nd}$} part)}}}
\`\fbox{{\bf{CIRCULARITY THM}}}
\end{tabbing}

\begin{tabbing}
\fbox{d.o.c.\ static and strictly stationary}
\`\fbox{d.o.c.\ circular}
\end{tabbing}
\vspace{-.6cm}
\begin{tabbing}
[$k^{\mu}k_{\mu} < 0$]
\` $\exists$ coordinates $t$, $\varphi$:
$k^{\mu} = \partial_{t}$ and\\
$\exists$ coordinate $t$:
$k^{\mu} = \partial_{t}$ is hypersurface orthogonal
\` $m^{\mu} = \partial_{\varphi}$ are hypersurface orthogonal
\end{tabbing}

\vspace{.35cm}
\begin{tabbing}
\fbox{{\bf{STATIC UNIQUENESS THM}}}
\`\fbox{{\bf{CIRCULAR UNIQUENESS THM}}}
\end{tabbing}
\vspace{-.6cm}
\begin{tabbing}
[originally by means of Israel's thm,
\` [originally by means of Robinson's thm,\\
later by the positive energy thm]
\` later by $\sigma$-model/harmonic map identities\\
\` or by inverse scattering techniques]
\end{tabbing}

\begin{tabbing}
\fbox{{\bf{Schwarzschild (Reissner--Nordstr\"om or MP)}}}
\`\fbox{{\bf{Kerr (Kerr--Newman) metric}}}
\end{tabbing}

\vspace{5mm}
\caption{Classification of \emph{analytic, connected, mean
    non-degenerate}, asymptotically-flat, $I^+$-regular, stationary
  electrovacuum black holes.}
\label{fig:classification}
\end{figure}
}

The hypotheses of analyticity and non-degeneracy are highly
unsatisfactory, and one believes that they are not needed for the
conclusion. One also believes that, in vacuum, the hypothesis of
connectedness is spurious, and that all black holes with more than
one component of the event horizon are singular, but no proof is
available except for some special cases~\cite{LiTian, Weinstein:trans,
  HennigNeugebauer3}. Indeed, Theorem~\ref{Tubh} should be compared
with the following conjecture, it being understood that both the
Minkowski and the Reissner--Nordstr\"om spacetimes are members of
the Kerr--Newman family:

\begin{Conjecture}
\label{Cubh}
Let $(\mcM,\fourg)$ be an electrovacuum, four-dimensional spacetime
containing a spacelike, connected, acausal hypersurface $\hyp $, such
that $\ohyp $ is a topological manifold with boundary, consisting of
the union of a compact set and of a finite number of asymptotically-flat ends. Suppose that there exists on $\mcM$ a complete stationary
Killing vector $\changedX$, that $\doc$ is globally hyperbolic, and
that $ \pohyp\subset\mcM\setminus \doc$. Then $\doc$ is isometric
to the domain of outer communications of a Kerr--Newman or
MP spacetime.
\end{Conjecture}

\subsubsection{A uniqueness theorem for near-Kerrian smooth vacuum stationary spacetimes}
\label{subsect-SmoothUT}

The existing results on \emph{rigidity without analyticity} require one to assume either staticity, or a near-Kerr condition on the spacetime geometry (see Section~\ref{S3VI11.3}), which is quantified in terms of a smallness condition of the Mars--Simon tensor~\cite{MarsKerr, SimonKerr}.
The results in~\cite{AIK2} together with
Theorems~\ref{Tubhelvstatic}\,--\,\ref{Tubhaxi}, and a version of the
R\'acz--Wald Theorem~\cite[Proposition 4.1]{FRW}, lead to:
\begin{theorem}
\label{TAIK}
Let $(\mcM,\fourg)$ be a stationary
asymptotically-flat, $I^+$-regular, smooth, vacuum four-dimensional
spacetime. Assume that the event horizon is connected and mean non-degenerate.
If the Mars--Simon tensor $S$ and the Ernst potential $\erpot$ of the spacetime satisfy
$$
\sum_{ijkl} |(1-\erpot)S_{ijkl}|<\epsilon
$$
for a small enough $\epsilon>0$, then $\doc$ is isometric to the
domain of outer communications of a Kerr spacetime.
\end{theorem}

%
%

\epubtkImage{classification2.png}{%
\begin{figure}[htbp]
\small
\begin{center}
\fbox{{\bf $(M,\fourg)$ Stationary, asymptotically flat, electrovacuum, $I^+$-regular black-hole spacetime}}\\
\vspace{.15cm}
[asymptotically time-like Killing field $\changedX$; classification of isometry groups and actions]
\end{center}

\vspace{.20cm}
\begin{center}
\fbox{{\bf{Regularity and topology theorems}} }\\
\vspace{.15cm}
[horizons smooth/analytic with spherical sections; d.o.c.\  simply connected; structure theorem; { nonexistence of prehorizons meeting the d.o.c.}]
\end{center}

\vspace{.30cm}
\begin{center}
\fbox{{\bf{STRONG RIGIDITY THM ({\boldmath$1^{\rm st}$} part)}} }
\\
\vspace{.20cm}{
Assume moreover that $(\mcM, \fourg)$ is either a) analytic,
 or b) vacuum and near-Kerrian with connected and mean non-degenerate event horizon $\left(\langle\kappa\rangle\neq 0\right)$.}
\end{center}

\begin{center}
\fbox{Event horizon $=$ Killing horizon $H[\KV]$}\\
\vspace{.15cm}
\end{center}

\begin{tabbing}
\fbox{$H[\KV]$ non-rotating: { $\fourg(\changedX,\changedX)|_{H(\KV)} \equiv 0$}} 
\`\fbox{$H[\KV]$ rotating:  {$\fourg(\changedX,\changedX)|_{H(\KV)} \not\equiv 0$}} 
\\
If non-degenerate ($\kappa\neq 0$):
\`
\\
$\bullet$ Horizon essentially bifurcate
\`
\\
$\bullet$ Existence of maximal hypersurfaces
\`
\end{tabbing}

\begin{tabbing}
\fbox{{\bf{STATICITY THM }}}
\`\fbox{{\bf{STRONG RIGIDITY THM ({\boldmath$2^{\rm nd}$} part)}}}
\end{tabbing}

\begin{tabbing}
\fbox{d.o.c.\ static}
\`\fbox{d.o.c.\ axisymmetric}
\end{tabbing}
\vspace{-.6cm}
\begin{tabbing}
$d\changedX^{\flat}\wedge \changedX^{\flat}=0$
\` If $(\mcM, \fourg)$ analytic or near-Kerrian:
\\
\` $\exists$ periodic Killing field $\myeta$ s.t.\ $[\changedX,\myeta]=0$
\\
\` $\mcM\approx\R\times\left(\R^3\setminus\cup_iB_i\right) $
\\
\`$\rho^2:=\fourg^2(\changedX,\myeta)-\fourg(\changedX,\changedX)\fourg(\myeta,\myeta)\geq 0$ in d.o.c.\
\end{tabbing}

\vspace{.20cm}
\begin{tabbing}
\fbox{{\bf{STATICITY THM ({\boldmath$2^{\rm nd}$} part)}}}
\`\fbox{{\bf{CIRCULARITY THM}}}
\end{tabbing}

\begin{tabbing}
\fbox{d.o.c.\ static and strictly stationary}
\`\fbox{d.o.c.\ circular}
\end{tabbing}
\vspace{-.6cm}
\begin{tabbing}
d.o.c.\  strictly stationary: $\fourg(\changedX,\changedX) < 0$

\` $\exists$ global Weyl coordinates $(t,\varphi,\rho,z)$:
\\
$\exists$ coordinate $t$: $\changedX = \partial_{t}$ is
\` $\changedX = \partial_{t}$, $\myeta = \partial_{\varphi}$
\\
hypersurface orthogonal
\` $\fourg=-\rho^2e^{2\llambda}dt^2+e^{-2\llambda}(d\varphi-vdt)^2+e^{2\, u}(d\rho^2+dz^2)$
\\
$\fourg=-V^2dt^2+\gamma_{ij}dx^idx^j$
\` (with controlled asymptotic behavior)
\end{tabbing}

\vspace{.30cm}
\begin{tabbing}
\fbox{{\bf{STATIC UNIQUENESS THM}}}
\`\fbox{{\bf{CIRCULAR UNIQUENESS THM}}}
\end{tabbing}
\vspace{-.6cm}
\begin{tabbing}
$\bullet$ originally by means of Israel's thm,
\` $\bullet$ originally by means of Robinson's thm,
\\
later by the positive energy thm
\` later by $\sigma$-model/harmonic map identities
\\

$\bullet$ No {analyticity}, degeneracy or
\` or by inverse scattering techniques
\\
  connectedness needed
\`
$\bullet$ {  multi-black holes?}
\end{tabbing}

\begin{tabbing}
\fbox{{\bf{Reissner--Nordstr\"om or MP}}}
\`\fbox{{\bf{Kerr--Newman metric}}}
\end{tabbing}

\caption{Classification of stationary electrovacuum black-hole spacetimes}
\label{fig:classification2}
\end{figure}}

%

\subsection{Summary of open problems}
\label{S13VI11.2}

For the convenience of the reader, we summarize here the main open
problems left in the no-hair theorem.

\subsubsection{Degenerate horizons}
\label{subsect-CLASS-BHDH}

We recall that there exist no vacuum \emph{static} spacetimes
containing \emph{degenerate} horizons with compact spherical
sections~\cite{CRT}. On the other hand,
MP~\cite{M47PM, P45PM} black holes provide the only
electro-vacuum static examples with non-connected degenerate horizons.
See~\cite{CN95PM, MH97PM} and references therein for a discussion of
the geometry of MP black holes.

Under the remaining hypotheses of Theorem~\ref{Tubh}, the only step
where the hypothesis of non-degeneracy enters is the proof of
existence of a maximal hypersurface $\hyp$ in the black-hole
spacetime under consideration, such that $\hyp$ is Cauchy for the
domain of outer communications. The geometry of Cauchy surfaces in the
case of degenerate horizons is well understood by
now~\cite{ChstaticelvacarxivErr, ChNguyen}, and has dramatically
different properties when compared to the non-degenerate case. A proof
of existence of maximal hypersurfaces in this case would solve the
problem, but requires new insights. A key missing element is an
equivalent of Bartnik's \emph{a priori} height
estimate~\cite{Bartnik84}, established for asymptotically-flat ends,
that would apply to asymptotically-cylindrical ends.

\subsubsection{Rigidity without analyticity}
\label{subsec-RigOpen}

Analyticity enters the current argument at two places: First, one
needs to construct the second Killing vector near the horizon. This
can be done by first constructing a candidate at the horizon, and then
using analyticity to extend the candidate to a neighborhood of the
horizon. Next, the Killing vector has to be extended to the whole
domain of outer communications. This can be done using analyticity and
a theorem by Nomizu~\cite{Nomizu}, together with the fact that
$I^+$-regular domains of outer communications are simply
connected. Finally, analyticity can be used to provide a simple
argument that prehorizons do not intersect $\doc$ (but this is not
critical, as a proof is available now within the smooth category of
metrics~\cite{ChGstatic}).

A partially-successful strategy to remove the analyticity condition
has been invented by Alexakis, Ionescu and Klainerman
in~\cite{AIK}. Their argument applies to non-degenerate near-Kerrian
configurations, but the general case remains open.

The key to the approach in~\cite{AIK} is a unique continuation theorem
near bifurcate Killing horizons proven in~\cite{IonescuKlainerman1},
which implies the existence of a second Killing vector field, say
$\myeta$, in a neighborhood of the horizon. One then needs to prove
that $\myeta$ extends to the whole domain of outer
communications. This is established via another unique continuation
theorem~\cite{IonescuKlainerman2} with specific convexity
conditions. These lead to non-trivial restrictions, and so far the
argument has only been shown to apply to near-Kerrian configurations.

A unique continuation theorem across more general timelike surfaces
would be needed to obtain the result without smallness restrictions.

It follows from what has been said in~\cite{ChGstatic} that the
boundary of the set where two Killing vector fields are defined cannot
become null within a domain of outer communications; this fact might
be helpful in solving the full problem.

\subsubsection{Many components?}
\label{ss14VI11.3}

The only known examples of singularity-free stationary electrovacuum
black holes with more than one component are provided by the
MP family. (Axisymmetric MP solutions are possible,
but MP metrics only have one Killing vector in general.) It has been
suspected for a very long time that these are the only such solutions,
and that there are thus no such vacuum configurations. This should be
contrasted with the five-dimensional case, where the Black Saturn
solutions of Elvang and Figueras~\cite{Elvang:2007rd}
(compare~\cite{CES, Szybka:Causality}) provide non-trivial
two-component examples.

It might be convenient to summarize the general facts known about
four-dimensional multi-component solutions.%
\epubtkFootnote{Here we are interested in stationary multi--black-hole
  configurations; nonexistence of some suitably regular stationary
  $n$-body configurations was established, under different symmetry
  conditions, in~\cite{Beig:2009jd, Beig:2008qi}.}
In case of doubts, $I^+$-regularity should be assumed.

We start by noting that the static solutions, whether connected or
not, have already been covered in Section~\ref{S3VI11.1}.

A multi-component electro-vacuum configuration with all components
non-degenerate and \emph{non-rotating} would be, by what has been
said, static, but then no such solutions exist (all components of an
MP black hole are degenerate). On the other hand, the question of
existence of a multi--black-hole configuration with components of
mixed type, none of which rotates, is open; what's missing is the
proof of existence of maximal hypersurfaces in such a case. Neither
axisymmetry nor staticity is known for such configurations.

Analytic multi--black-hole solutions with at least one rotating
component are necessarily axisymmetric; this leads one to study the
corresponding harmonic-map equations, with candidate solutions
provided by Weinstein or by inverse scattering
techniques~\cite{KramerNeugebauerDoubleKerr, Weinstein3,
  BelinskiZakharov, BelinskiZakharov2}. The Weinstein solutions have
no singularities away from the axes, but they are not known in
explicit form, which makes difficult the analysis of their behavior
on the axis of rotation. The multi--black-hole metrics constructed by
multi-soliton superpositions or by B\"acklund transformation
techniques are obtained as rational functions with several parameters,
with explicit constraints on the parameters that lead to a regular
axis~\cite{MankoRuizSanabria}, but the analysis of the zeros of their
denominators has proved intractable so far. It is perplexing that the
five dimensional solutions, which are constructed by similar
methods~\cite{Pomeransky}, can be completely analyzed with some effort
and lead to regular solutions for some choices of parameters, but the
four-dimensional case remains to be understood.

In any case, according to Varzugin~\cite{Varzugin1, Varzugin2} and,
independently, to Neugebauer and Meinel~\cite{Neugebauer:2003qe} (a
more detailed exposition can be found in~\cite{HennigNeugebauer3,
  HennigNeugebauer2}), the multi-soliton solutions provide the only
candidates for stationary axisymmetric electrovacuum solutions. A
breakthrough in the understanding of vacuum \emph{two-component}
configurations has been made by Hennig and
Neugebauer~\cite{HennigNeugebauer2, HennigNeugebauer3}, based on the
area-angular momentum inequalities of Ansorg, Cederbaum and
Hennig~\cite{HennigAnsorgCederbaum} as follows: Hennig and Neugebauer
exclude many of the solutions by verifying that they have negative
total ADM mass. Next, configurations where two horizons have vanishing
surface gravity are shown to have zeros in the denominators of some
geometric invariants. For the remaining ones, the authors impose a
non-degeneracy condition introduced by Booth and
Fairhurst~\cite{BoothFairhurst}: a black hole is said to be
\emph{sub-extremal} if any neighborhood of the event horizon contains
trapped surfaces. The key of the analysis is the angular momentum -
area inequality of Hennig, Ansorg, and
Cederbaum~\cite{HennigAnsorgCederbaum}, that on every {sub-extremal}
component of the horizon it holds that
\begin{equation}
\label{14VI11.10}
 8 \pi |
J| < A
 \,,
\end{equation}
where $J$ is the Komar angular-momentum and $A$ the area of a
section. (It is shown in~\cite{AnsorgPfister} and
\cite[Appendix]{HCAUniversalInequality} that $\kappa=0$ leads to
equality in \eq{14VI11.10} under conditions relevant to the problem at
hand.) Hennig and Neugebauer show that all remaining candidate
solutions violate the inequality; this is their precise non-existence
statement.

The problem with the argument so far is the lack of justification of
the sub-extremality condition. Fortunately, this condition can be
avoided altogether using ideas of~\cite{DainReiris} concerning the
inequality~\eqref{14VI11.10} and appealing to the results
in~\cite{Eichmair, AndM2, CGS} concerning \emph{marginally--outer-trapped surfaces} (MOTS): Using existence results of weakly stable
MOTS together with various aspects of the candidate Weyl metrics, one
can adapt the argument of~\cite{HennigAnsorgCederbaum} to
show~\cite{CENS} that the area inequality~\eq{14VI11.10}, with ``less
than'' there replaced by ``less than or equal to'', would hold for
those components of the horizon, which have non-zero surface gravity,
assuming an $I^+$-regular metric of the Weyl form, if any existed. The
calculations of Hennig and Neugebauer~\cite{HennigNeugebauer3}
together with a contradiction argument lead then to
\begin{Theorem}
 \label{T26VII11}
 $I^+$-regular two-Kerr black holes do not exist.
\end{Theorem}

The case of more than two horizons is widely open.

\newpage

\section{Classification of stationary toroidal Kaluza--Klein black holes}
\label{sec-KK}

In this work we are mostly interested in uniqueness results for
four-dimensional black holes. This leads us naturally to consider
those vacuum Kaluza--Klein spacetimes with enough symmetries to lead
to four-dimensional spacetimes after dimensional reduction, providing
henceforth four-dimensional black holes. It is convenient to start
with a very short overview of the subject; the reader is referred to
\cite{EmparanReallLR, HorowitzWiseman} and references therein for more
information. Standard examples of Kaluza--Klein black holes are
provided by the Schwarzschild metric multiplied by any spatially flat
homogeneous space (e.g., a torus). Non-trivial examples can be found
in \cite{Rasheed, LarsenKK}; see also~\cite{KudohWiseman,
  HorowitzWiseman} and reference therein.

\subsection{Black holes in higher dimensions}
\label{sect-Hdim}

The study of spacetimes with dimension greater then four is almost as
old as general relativity itself~\cite{Kaluza:1921tu,
  Klein:1926tv}. Concerning black holes, while in dimension four all
explicitly-known asymptomatically-flat and regular solutions of the
vacuum Einstein equations are exhausted by the Kerr family, in
spacetime dimension five the landscape of known solutions is
richer. The following $I^+$-regular, stationary, asymptotically-flat,
vacuum solutions are known in closed form: the Myers--Perry black
holes, which are higher-dimensional generalizations of the Kerr metric
with spherical-horizon topology~\cite{Myers:1986un}; the celebrated
Emparan--Reall black rings with $S^2\times S^1$ horizon
topology~\cite{EmparanReall}; the Pomeransky--Senkov black rings
generalizing the previous by allowing for a second angular-momentum
parameter~\cite{PS}; and the ``Black Saturn'' solutions discovered by
Elvang and Figueras, which provide examples of regular multi-component
black holes where a spherical horizon is surrounded by a black
ring~\cite{Elvang:2007rd}.%
\epubtkFootnote{Studies of regularity and causal structure of black
  rings and Saturns can be found in~\cite{CSPS, CC, CCGP,
    Szybka:Causality, CES}.}

Inspection of the basic features of these solutions challenges any
naive attempt to generalize the classification scheme developed for
spacetime dimension four: One can find black rings and Myers--Perry
black holes with the same mass and angular momentum, which must
necessarily fail to be isometric since the horizon topologies do not
coincide. In fact there are non-isometric black rings with the same
Poincar\'e charges; consequently a classification in terms of mass,
angular momenta and horizon topology also fails. Moreover, the Black
Saturns provide examples of regular vacuum multi--black-hole solutions,
which are widely believed not to exist in dimension four;
interestingly, there exist Black Saturns with vanishing total angular
momentum, a feature that presumably distinguishes the Schwarzschild
metric in four dimensions.

Nonetheless, results concerning 4-dimensional black holes either
generalize or serve as inspiration in higher dimensions. This is true
for landmark results concerning black-hole uniqueness and, in fact,
classification schemes exist for classes of higher dimensional
black-hole spacetimes, which mimic the symmetry properties of the
``static or axisymmetric'' alternative, upon which the uniqueness
theory in four-dimensions is built.

For instance, staticity of $I^+$-regular, vacuum, asymptotically-flat,
non-rotating,  non-degenerate black holes remains true in higher
dimensions%
\epubtkFootnote{It should be noted that, although formulated for
  4-dimensional spacetimes, the results in~\cite{CW94MAX} remain
  valid without changes in higher-dimensional spacetimes.}.
Also, Theorem~\ref{Tubhelvstatic} remains valid for vacuum spacetimes
of dimension $n+1$, $n\geq3$, whenever the positive energy theorem
applies to an appropriate doubling of $\hyp $ (see~\cite{ChGstatic},
Section~\ref{S3VI11.1} and references therein). Moreover, the
discussion in Section~\ref{S3VI11.1} together with the results
in~\cite{Rogatko:2003kj, Rogatko:2006gg} suggest that an analogous
generalization to electrovacuum spacetimes exists, which would lead to
uniqueness of the higher-dimensional Reissner--Nordstr\"om metrics
within the class of static solutions of the Einstein--Maxwell
equations, for all $n\geq 3$ (see also~\cite[Section
  8.2]{EmparanReallLR}, \cite{Ida:2011jm} and references therein).

Rigidity theorems are also available for $(n+1)$-dimensional,
asymptotically-flat and analytic black-hole spacetimes: the
non-degenerate horizon case was established in~\cite{HIW}
(compare~\cite{Moncrief:2008mr}), and partial results concerning the
degenerate case were obtained in~\cite{Hollands:2008wn}. These show
that stationary rotating (analytic) black holes are ``axisymmetric'',
in the sense that their isometry group contains $\R\times U(1)$; the
$\R$ factor corresponds to the action generated by the stationary
vector, while the circle action provides an ``extra'' axial Killing
vector. A conjecture of Reall~\cite{Reall:2002bh}, supported by the
results in~\cite{Emparan:2009vd}, predicts the existence of
5-dimensional black holes with exactly $\R\times U(1)$ isometry group;
in particular, it is conceivable that the rigidity results are sharp
when providing only one ``axial'' Killing vector. The results
in~\cite{Gubser, DFMRS, HollandsWaldStability} are likely to be
relevant in this context.

So we see that, assuming analyticity and asymptotic flatness, the
dichotomy provided by the rigidity theorem remains valid but its
consequences appear to be weaker in higher dimensions. A gap appears
between the two favorable situations encountered in dimension four:
one being the already discussed staticity and the other corresponding
to black holes with cohomogeneity-two Abelian groups of isometries. We
will now consider this last scenario, which turns out to have
connections to the four dimensional case.

\subsection{Stationary toroidal Kaluza--Klein black holes}
\label{subsec-KK}

The four-dimensional vacuum Einstein equations simplify considerably
in the stationary and axisymmetric setting by reducing to a harmonic
map into the hyperbolic plane (see Sections~\ref{sect-SAST}
and~\ref{sec-red}). A similar such reduction in $(n+1)$-dimensions
works when the isometry group includes $\R\times \T^{n-2}$, i.e.,
besides the stationary vector there exist $n-2$ commuting axial
Killing vectors.

Since the center $\T^s$ of $SO(n)$ has dimension
$$
s=\left\lfloor\frac{n}{2}\right\rfloor \,,
$$
in the asymptotically flat case the existence of such a group of
isometries is only possible for $n=3$ or $n=4$. However, one can move
beyond the usual asymptotic-flatness and consider instead
$KK$-asymptotically-flat spacetimes, in the sense of
Section~\ref{subsec-KKdefxx}, with asymptotic ends $\hyp_{\ext}\approx
(\R^{\Ldim}\setminus B)\times\T^s$, satisfying ${\Ldim}=3,4$ and
${\Ldim}+s=n$, with the isometry group containing $\R\times\T^{n-2}$,
$n\ge 3$. Here one takes $\Mtext\approx\left(\R^{\Ldim}\setminus\bar
B(R)\right)\times \T^s$, with the reference metric of the form
$\mathring g=\delta\oplus\mathring h$, where $\mathring h$ is the flat
$s$-torus metric. Finally, the action of $\R\times \T^{n-2}$ on
$(\mcM,\fourgx)$ by isometries is assumed in the exterior region
$\Mext\approx\R\times\Mtext$ to take the form
\begin{equation}
\label{KKaction}
  \R\times \T^{n-2}\ni (\tau, g): \quad\quad (t,p)\mapsto (t+\tau,g\cdot p)\,.
\end{equation}
Such metrics will be referred to as \emph{stationary toroidal
  Kaluza--Klein metrics}.

\subsection{Topology of the event horizon}
\label{subsec-TopEH}

A theorem of Galloway and Schoen~\cite{Galloway:2005mf} shows that
compact cross-sections of the horizon must be of positive Yamabe type,
i.e., admit metrics of positive scalar curvature. In spacetime
dimension five, the positive Yamabe property restricts the horizon to
be a finite connected sum of spaces with $S^3$, $S^2\times S^1$ and
lens-space $L(p,q)$ topologies. Such results require no symmetry
assumptions but by further assuming stationarity and the existence of
one axial Killing vector more topological restrictions, concerning the
allowed factors in the connected sum, appear in five
dimensions~\cite{Hollands:2010qy}.

In the toroidal Kaluza--Klein case, the existence of a toroidal action
leads to further restrictions~\cite{Hollands:2008fm}; for instance,
for ${\Ldim}=4$ and $n\geq 4$, each connected component of the horizon
has necessarily one of the following topologies: $S^3\times \T^{n-4}$,
$S^2\times T^{n-3}$ and $L(p,q)\times \T^{n-4}$.

It should be noted that no asymptotically-flat or Kaluza--Klein black
holes with lens-space topology of the horizon are known. Constructing
a black lens, or establishing non-existence, appears to be a
challenging problem.

\subsection{Orbit space structure}
\label{subsec-OrbitSpace}

The structure theorem~\cite{ChHighDim} applies to stationary toroidal
Kaluza--Klein black holes and provides the following product structure
$$
{\doc}\approx \R\times \Sigma \,,
$$
with the stationary vector being tangent to the $\R$ factor and where
$\Sigma$, endowed with the induced metric, is an $n$-dimensional
Riemannian manifold admitting a $\T^{n-2}$ action by isometries. A
careful analysis of the topological properties of such toroidal
actions~\cite{Hollands:2008fm}, based on deep results
from~\cite{OrlikRaymondI, OrlikRaymondII}, allows one to show that the
orbit space $\doc/(\R\times \T^{n-2})$ is homeomorphic to a half plane
with boundary composed of segments and corners; the segments being the
projection of either a component of the event horizon or an axis of
rotation (the set of zeros of a linear combination of axial vectors),
and the corners being the projections of the intersections of two
axes. Moreover, the interiors are in fact diffeomorphic. To establish
this last fundamental result it is necessary to exclude the existence
of exceptional orbits of the toroidal action; this was done by
Hollands and Yazadjiev in~\cite{Hollands:2008fm} by extending the
results in~\cite{OrlikRaymondI} to the $KK$-black hole setting. In
particular one obtains the following decomposition
$$
{\doc}\setminus(\cup\mcA_i)\approx \R\times \T^{n-2}\times\R\times\R^+ \,,
$$
where $\cup\mcA_i$ is the union of all axes; we note that such product
structure is necessary to the construction of Weyl
coordinates~\cite{ChCo, ChHighDim} and, consequently, indispensable to
perform the desired reduction of the vacuum equations.

As already discussed, basic properties of black rings show that a
classification of $KK$-black holes in terms of mass, angular momenta
and horizon topology is not possible. But, as argued by Hollands and
Yazadjiev~\cite{Hollands:2008fm}, the angular momenta and the
structure of the orbit space characterize such black holes if one
further assumes non-degeneracy of the event horizon. This orbit space
structure is in turn determined by the \emph{interval structure} of
the boundary of the quotient manifold, a concept related to Harmark's
\emph{rod structure} developed in~\cite{Harmark:2004rm} (see
also~\cite[Section 5.2.2.1]{EmparanReallLR}). Note that the interval
structure codifies the horizon topology.

\subsection{KK topological censorship}
\label{subsec-topcKK}

Black-hole uniqueness in four-dimensions uses simple connectedness of
the event horizon extensively. But the Schwarzschild metric multiplied
by a flat torus shows that simple connectedness does not hold for
general domains of outer communications of Kaluza--Klein black
holes. Fortunately, simple connectedness of the orbit space
$\doc/(\R\times \T^{n-2})$ suffices: for instance, to prove that the
$(n-1)$-dimensional orbit generated by the stationary and axial
vectors is timelike in $\doc$ away from the axes (which in turn is
essential to the construction of Weyl coordinates), to guarantee the
existence of global twist potentials~\cite{ChHighDim} and to exclude
the existence of exceptional orbits of the toroidal action (see
Section~\ref{subsec-OrbitSpace}). The generalized topological
censorship theorem of~\cite{CGS} shows that this property follows from
the simple connectedness of the orbit space in the asymptotic end
$\Mtext/\T^{n-2}\approx\R^{{\Ldim}}\setminus B$.

\subsection{Classification theorems for \textit{KK}-black holes}
\label{subsec-ClssKK}

As usual, the static case requires separate consideration. The first
classification results addressed static five-dimensional solutions
with KK asymptotics and with a $\R\times U(1)$ factor in the group of
isometries. In such a setting, the Kaluza--Klein reduction leads to
gravity coupled with a Maxwell field $F$ and a ``dilaton'' field
$\phi$, with a Lagrangean
$$
L = R - 2 (\nabla \phi)^2 - e^{-2\alpha \phi} |F|^2 \,,
$$
where $\alpha=\sqrt 3$. In the literature one also considers more
general theories where $\alpha$ does not necessarily take the
Kaluza--Klein value. All current uniqueness proofs require that the
mass, the Maxwell charges, and the dilaton charge satisfy a certain
\emph{genericity condition}, and that all horizon components have
non-vanishing surface gravity. When $\alpha=1$, Mars and
Simon~\cite{MarsSimonDilaton} show that the generic static solutions
belong to the family found by Gibbons and Maeda~\cite{GWGM88,
  GWG82dil, GHS91}. For other values of $\alpha$, in particular for
the KK value, a \emph{purely electric} or \emph{purely magnetic}
configuration is assumed, and then the same conclusion is reached. The
result is an improvement on the original uniqueness theorems of
Simon~\cite{WS85UT} and Masood-ul-Alam~\cite{MuA93dil}, and has
been generalized to higher dimensions in~\cite{GIS2}. The analyticity
assumption, which is implicit in all the above proofs, can be removed
using~\cite{ChGstatic}.

The remaining classfication results assume cohomogeneity-two isometry
actions~\cite{Hollands:2008fm}:

\begin{theorem}
\label{TubhaxiKK}
Let $(\mcM_i,\fourg_i)$, $i=1,2$, be two $I^+$-regular,
$(n+1)$-dimensional, $n\geq 3$, stationary toroidal Kaluza--Klein
spacetimes, with five asymptotically-large dimensions
(${\Ldim}=4$). Assume, moreover, that the event horizon is connected and
mean non-degenerate. If the interval structure and the set of angular
momenta coincide, then the domains of outer communications are
isometric.
\end{theorem}

This theorem generalizes previous results by the same
authors~\cite{HY, HY2} as well as a uniqueness result for a connected
spherical black hole of~\cite{Morisawa:2004tc}.

The proof of Theorem~\ref{TubhaxiKK} can be outlined as follows: After
establishing the, mainly topological, results of Sections~\ref{subsec-TopEH},~\ref{subsec-OrbitSpace} and~\ref{subsec-topcKK}, the proof follows closely the arguments for uniqueness of
4-dimensional stationary and axisymmetric electrovacuum black
holes. First, a generalized Mazur identity is valid in higher
dimensions (see~\cite{Maison:Ehlers, BMG88KK} and
Section~\ref{subsec-TMI}). From this Hollands and Yazadjiev show that
(compare the \newcommand{\HYsigma}{\psi} discussion in
Sections~\ref{subsec-DivId} and~\ref{ss14VI11.1})
\begin{equation}
\label{sigmaSubHarm}
\Delta_\delta \HYsigma \geq 0\,,
\end{equation}
where $\Delta_\delta$ is the flat Laplacian on $\R^3$, the function
$$
\HYsigma:\R^3\setminus\{z=0\}\rightarrow \R\,,
$$
is defined as
$$
\HYsigma=\mbox{Trace}(\Phi_{2} \Phi_{1}^{-1} - \mathbb{1})\,,
$$
the $\Phi_i$'s, $i=1,2$, are the Mazur
matrices~\cite[Eq.~(78)]{Hollands:2008fm} associated with the two
black-hole spacetimes that are being compared. In terms of twist
potentials $(^{(i)}\chi_i)$ and metric components of the axisymmetric
Killing vectors (generators of the toroidal symmetry)
$$
 ^{(i)}f_{mn}=\left(^{(i)}\fourg(k_m,k_n)\right)\,,\;\;i=1,2\,,
$$
we have the following explicit formula (see also~\cite{Maison:Ehlers,
  Morisawa:2004tc})
\begin{equation}
\label{sigma}
\HYsigma=-1+\frac{^{(1)}f}{^{(2)} f}+\frac{^{(1)}f^{ij}\left(^{(1)}\chi_i-^{(2)}\chi_i\right)
\left(^{(1)}\chi_j-^{(2)}\chi_j\right)}{^{(2)} f}
+\, ^{(1)}f^{ij}\left( ^{(2)} f_{ij}-\,^{(1)}f_{ij}\right)
 \,,
\end{equation}
where $^{(i)}f=\det\left(^{(i)}f_{mn}\right),$ and where
$^{(1)}f^{ij}$ is the matrix inverse to $^{(1)}f_{ij}$.

It should be noted that this provides a variation on Mazur's and the
harmonic map methods (see Sections~\ref{ss11VI11.1}
and~\ref{ss11VI11.2}), which avoids some of their intrinsic
difficulties. Indeed, the integration by parts argument based on the
Mazur identity requires detailed knowledge of the maps under
consideration at the singular set $\{\rho =0\}$, while the harmonic
map approach requires finding, and controlling, the distance function
for the target manifold. (In some simple cases $\HYsigma$ is the
desired distance function, but whether this is so in general is
unclear.) The result then follows by a careful analysis of the
asymptotic behavior of the relevant fields; such analysis was also
carried out in~\cite{Hollands:2008fm}.

In this context, the degenerate horizons suffer from the supplementary
difficulty of controlling the behavior of the fields near the
horizon. One expects that an exhaustive analysis of near-horizon
geometries would allow one to settle the question; some partial
results towards this can be found in~\cite{KunduriLuciettiReall,
  KunduriLucietti, KunduriLucietti2, HollandsIshibashiKK}.

\newpage

\section{Beyond Einstein--Maxwell}
\label{sect-BEM}

The purpose of this section is to reexamine the various steps leading
to the classification of electrovacuum black-hole spacetimes for
other matter models. In particular, it will be seen that several steps
in Figure~\ref{fig:classification} cease to hold in the presence of
non-Abelian gauge fields. Unfortunately, this implies that we are far
from having a classification of all stationary black-hole spacetimes
with physically-interesting sources.

\subsection{Spherically symmetric black holes with hair}
\label{subsect-BEM-SSBHH}

One can find in the literature the naive expectation that -- within a
\textit{given} matter model -- the stationary black-hole solutions are
uniquely characterized by a set of global charges; this will be
referred to as the \emph{generalized no-hair conjecture}. A model in
which this might possibly be correct is provided by the static sector
of the EM-dilaton theory, discussed at the beginning of
Section~\ref{subsec-ClssKK}.

The failure of this generalized no-hair conjecture is demonstrated by
the Einstein--Yang--Mills (EYM) theory: According to the conjecture,
any static solution of the EYM equations should either coincide with
the Schwarzschild metric or have some non-vanishing Yang--Mills
charges. This turned out not to be the case when, in 1989, various
authors~\cite{VG89firsthair, KMuA90firsthair, PB90firsthair} found a
family of static black-hole solutions with vanishing global
Yang--Mills charges (as defined, e.g., in~\cite{ChK}); these were
originally constructed by numerical means and rigorous existence
proofs were given later in~\cite{SWY91sol, SW93sol, SW93bh, BFM94EX,
  MavromatosWinstanley}; for a review see~\cite{VolkovGaltsov}. These
solutions violate the generalized no-hair conjecture.

As the non-Abelian black holes are unstable~\cite{SZ90Stab, ZS91Stab,
  BW92Stab}, one might adopt the view that they do not present actual
threats to the generalized no-hair conjecture. (The reader is referred
to~\cite{BHS96Puls} for the general structure of the pulsation
equations, \cite{VBLS95Stab, BS94Stab}, to~\cite{BB94Zh} for the
sphaleron instabilities of the particle-like solutions, and
to~\cite{SP93Rev} for a review on sphalerons.) However, various authors
have found stable black holes, which are not characterized by a set of
asymptotic flux integrals. For instance, there exist stable black-hole
solutions with hair of the static, spherically-symmetric
Einstein--Skyrme equations~\cite{DHS91Skyrme, HDS91Skyrme,
  HDS92Skyrme, HSZ93Skyrme, MossShikiWinstanley} and to the EYM
equations coupled to a Higgs triplet~\cite{BFM92monop, BFM95monop2,
  EW94Top, AB93}; it should be noted that the solutions of the
EYM--Higgs equations with a Higgs doublet are unstable~\cite{BB94Zh,
  WM95}. Hence, the restriction of the generalized no-hair conjecture
to stable configurations is not correct either.

One of the reasons why it was not until 1989 that black-hole
solutions with self-gravitating gauge fields were discovered was
the widespread belief that the EYM equations admit \textit{no soliton}
solutions. There were, at least, five reasons in support of
this hypothesis.

\begin{itemize}

\item First, there exist \textit{no purely gravitational solitons},
  that is, the only globally-regular, asymptotically-flat, static
  vacuum solution to the Einstein equations with finite energy is
  Minkowski spacetime. This is the \emph{Lichnerowicz theorem}, which
  nowadays can be obtained from the positive mass theorem and the Komar
  expression for the total mass of an asymptotically-flat, stationary
  spacetime~\cite{Horowitz}; see, e.g.,~\cite{GWG90SMM}
  or~\cite{MH96HPA}. A rather strong version thereof, which does not
  require asymptotic conditions other than completeness of the space
  metric, has been established by Anderson~\cite{manderson:static},
  see also~\cite{manderson:stationary}.

\item Next, there are no nontrivial static solutions of the EYM
  equations near Minkowski spacetime~\cite{MalecYM}.

\item Further, both Deser's energy argument~\cite{SD76} and
Coleman's scaling method~\cite{Col77} show that there do not exist
\textit{pure YM solitons} in flat spacetime.

\item Moreover, the EM system admits \textit{no soliton} solutions.
(This follows by applying Stokes' theorem to the
static Maxwell equations; see, e.g.,~\cite{MH96LN}.)

\item Finally, Deser~\cite{SD84} proved that the \textit{three-dimensional}
EYM equations admit \textit{no soliton solutions}.
The argument takes advantage of the fact that
the magnetic part of the Yang--Mills field has only one non-vanishing
component in 2+1 dimensions.
\end{itemize}

All this shows that it was conceivable to conjecture a
nonexistence theorem for soliton solutions of the EYM
equations in 3+1 dimensions, and a no-hair theorem
for the corresponding black hole configurations.
On the other hand, none of the above examples takes care of
the full nonlinear EYM system, which bears the possibility to
balance the gravitational and the gauge field interactions.
In fact, a closer look at the structure of the EYM
action in the presence of a Killing symmetry dashes the
hope to generalize the uniqueness proof
along the lines used in the Abelian case: The Mazur
identity owes its existence to the $\sigma$-model
formulation of the EM equations. The latter is, in turn,
based on \textit{scalar magnetic potentials}, the existence of which
is a peculiarity of Abelian gauge fields
(see Section~\ref{sec-SST}).

\subsection{Static black holes without spherical symmetry}
\label{subsect-BEM-SBH}

The above counterexamples to the generalized no-hair conjecture
are static \textit{and} spherically symmetric. The famous Israel
theorem guarantees that spherical symmetry is, in fact, a
consequence of staticity, provided that one is dealing with
vacuum~\cite{WI67vac} or electrovacuum~\cite{WI68evac} black-hole
spacetimes. The task to extend the Israel theorem to more
general self-gravitating matter models is, of course, a difficult
one. In fact, the following example proves that spherical
symmetry is not a generic property of static black holes.

In \cite{EW92RN}, Lee et al. reanalyzed
the stability of the Reissner--Nordstr\"om (RN) solution in the context
of $SU(2)$ EYM--Higgs theory. It turned out that -- for sufficiently
small horizons -- the RN black holes develop
an instability against radial perturbations of the Yang--Mills
field. This suggested the existence of magnetically-charged,
spherically-symmetric black holes with hair, which were also found
by numerical means~\cite{BFM92monop, BFM95monop2, EW94Top, AB93}.

Motivated by these solutions, Ridgway and
Weinberg~\cite{RW95nonsph} considered the stability of
the magnetically charged RN black holes within
a related model; the EM system coupled to a \textit{charged, massive
vector field}. Again, the RN solution
turned out to be unstable with respect to fluctuations of the
massive vector field. However, a perturbation analysis
in terms of spherical harmonics revealed that the fluctuations
cannot be radial (unless the magnetic charge assumes an
integer value), as discussed in Weinberg's comprehensive review on
magnetically-charged black holes~\cite{EW95MCBH}.
In fact, the work of Ridgway and Weinberg shows that
static black holes with magnetic charge need not even be
axially symmetric~\cite{RW95nonax}. Axisymmetric,
static black holes without spherical
symmetry appear to exist within the pure EYM system and the EYM-dilaton
model~\cite{BKJK97RB}.

This shows that static black holes may have considerably more
structure than one might expect from the experience with the
EM system: Depending on the matter model, they
may allow for nontrivial fields outside the horizon and,
moreover, they need not be spherically symmetric.
Even more surprisingly, there exist static black holes
without any rotational symmetry at all.

\subsection{The Birkhoff theorem}
\label{subsect-BEM-BT}

The Birkhoff theorem shows that the domain of outer communication of a
spherically-symmetric black-hole solution to the vacuum or the EM
equations is static. The result does not apply to many other matter
models: dust, fluids, scalar fields, Einstein-Vlasov, etc., and it is
natural to raise the question for non-Abelian gauge fields. Now, the
Einstein Yang--Mills equations have a well-posed Cauchy problem, so
one needs to make sure that the constraint equations admit
non-stationary solutions: Bartnik~\cite{BartnikHungary} has indeed
proved existence of such initial data. The problem has also been
addressed numerically in~\cite{Z92PhD, ZS91Stab}, where spherically-symmetric solutions of the EYM equations describing the explosion of a
gauge boson star or its collapse to a Schwarzschild black hole have
been found. A systematic study of the problem for the EYM system with
arbitrary gauge groups was performed by Brodbeck and
Straumann~\cite{BS93Birk}. Extending previous results of
K\"unzle~\cite{HPK91CQG} (see also~\cite{HPK94CMP, HPK94DG}), the
authors of~\cite{BS93Birk} were able to classify the principal bundles
over spacetime, which -- for a given gauge group -- admit $SO(3)$ as
symmetry group, acting by bundle automorphisms. It turns out that the
Birkhoff theorem can be generalized to bundles, which admit only
$SO(3)$--invariant connections of Abelian type. {We refer
  to~\cite{BS93Birk} for the precise formulation of the statement in
  terms of Stiefel diagrams, and to~\cite{OB95PHD, OB96HPA, HSV84,
    OliynykKunzleAll} for a classification of EYM solitons.} The
results in~\cite{OliynykFisher, BartnikFisherOliynyk} concerning
particle-like EYM solutions are likely to be relevant for the
corresponding black-hole problem, but no detailed studies of this
exist so far.

\subsection{The staticity problem}
\label{subsect-BEM-SP}

Going back one step further in the left half of the classification
scheme displayed in Figure~\ref{fig:classification}, one is led to the
question of whether all black holes with non-rotating horizon are
static. For non-degenerate EM black holes this issue was settled by
Sudarsky and Wald~\cite{SW92stat, SW93stat, CW94MAX},%
\epubtkFootnote{An early apparent success rested on a sign
  error~\cite{BC73BH}. Carter's amended version of the proof was
  subject to a certain inequality between the electric and the
  gravitational potential~\cite{BC87CAR}.}
while the corresponding vacuum problem was solved quite some time
ago~\cite{HE73LSS}; the degenerate case remains open. Using a slightly
improved version of the argument given in~\cite{HE73LSS}, the
staticity theorem can be generalized to self-gravitating stationary
scalar fields and scalar mappings~\cite{MH96HPA} as, for instance, the
Einstein--Skyrme system. (See also~\cite{MHNS93CQG, MH93Stat,
  MHNS94IJMP}, for more information on the staticity problem). It
should also be noted that the proof given in~\cite{MH96HPA} works
under less restrictive topological assumptions, since it does not
require the global existence of a twist potential.

While the vacuum and the scalar staticity theorems
are based on differential identities and integration by parts,
the approach due to Sudarsky and Wald takes advantage of
the ADM formalism and the existence of a maximal slicing~\cite{CW94MAX}.
Along these lines, the authors of~\cite{SW92stat, SW93stat}
were able to extend the staticity theorem to
topologically-trivial non-Abelian black-hole solutions. However, in contrast to the
Abelian case, the non-Abelian version applies only to configurations
for which either all components of the electric Yang--Mills charge
or the electric potential vanish asymptotically. This leaves some
room for stationary black holes, which are non-rotating and not
static. Moreover, the theorem implies that such configurations
must be charged. On a perturbative level, the existence of
these charged, non-static black holes with vanishing total angular
momentum was established in~\cite{BHSV97}.

\subsection{Rotating black holes with hair}
\label{subsect-BEM-RBH}

So far we have addressed the ramifications occurring on
the ``non-rotating half'' of the
classification diagram of Figure~\ref{fig:classification}:
We have argued that non-rotating black holes need not be static;
static ones need not be spherically symmetric; and spherically-symmetric ones need not be characterized by a set of global
charges. The right-hand-side of the classification scheme has been
studied less intensively so far. Here, the obvious questions
are the following:
Are all stationary black holes with rotating
Killing horizons axisymmetric (rigidity)?
Are the stationary and axisymmetric Killing fields
orthogonally-transitive (circularity)?
Are the circular black holes characterized by their mass, angular
momentum and global charges (no-hair)?

Let us start with the first issue, concerning the generality of the
strong rigidity theorem (SRT). The existence of a second Killing
vector field to the future of a bifurcation surface can be established
by solving a characteristic Cauchy problem~\cite{FRW}, which makes it
clear that axial symmetry will hold for a large class of matter models
satisfying the, say, dominant energy condition.

The counterpart to the staticity problem is the \textit{circularity}
problem: As general non-rotating black holes are not static,
one expects that the axisymmetric ones need not
be circular. This is, indeed, the case: While circularity is a
consequence of the EM equations and the symmetry properties
of the electro-magnetic field, the same is not true for the EYM
system. In the Abelian case, the proof rests on the
fact that the field tensor satisfies $F(k,m) = (\fourast F)(k,m) = 0$,
$k$ and $m$ being the stationary and the axial Killing field, respectively;
for Yang--Mills fields these conditions do no longer follow
from the field equations and their invariance properties
(see Section~\ref{subsec-CIR} for details). Hence,
the familiar Papapetrou ansatz for a stationary and axisymmetric
metric is too restrictive to take care of all stationary and
axisymmetric degrees of freedom of the EYM system. However, there
are other matter models for which the Papapetrou metric is
sufficiently general: the proof of the circularity theorem
for self-gravitating scalar fields is, for instance,
straightforward~\cite{MH95UROT}.
Recalling the key simplifications
of the EM equations arising from the (2+2)-splitting of the metric in
the Abelian case, an investigation of non-circular
EYM equations is expected to be rather awkward.
As rotating black holes with hair are most likely to
occur already in the circular sector (see the next paragraph),
a systematic investigation
of the EYM equations with circular constraints is needed as well.

The \textit{static} subclass of the circular sector was
investigated in studies by Kleihaus and Kunz
(see~\cite{BKJK97RB} for a compilation of the results).
Since, in general, staticity does not imply spherical symmetry,
there is a possibility for a static branch of axisymmetric
black holes without spherical symmetry.
Using numerical methods, Kleihaus and Kunz have constructed
black-hole solutions of this kind for both the EYM and the
EYM-dilaton system~\cite{BKJK97BH1}. The related
axisymmetric \textit{soliton} solutions without spherical
symmetry were previously obtained by the same
authors~\cite{BKJK97Sol1, BKJK97Sol2}; see
also~\cite{BKJK97Sol3} for more details. The new configurations are
purely magnetic and parameterized by their winding number and the node
number of the relevant gauge field amplitude. In the formal limit of
infinite node number, the EYM black holes approach the
Reissner--Nordstr\"om solution, while the EYM-dilaton black holes tend
to the Gibbons--Maeda black hole~\cite{GWG82dil, GWGM88}. The
solutions themselves are neutral and not spherically symmetric;
however, their limiting configurations are charged and spherically
symmetric. Both the soliton and the black-hole solutions of Kleihaus
and Kunz are unstable and may, therefore, be regarded as gravitating
sphalerons and black holes inside sphalerons, respectively.

Existence of \textit{slowly rotating} regular black-hole solutions to the
EYM equations was established in~\cite{BHSV97}.
Using the reduction of the EYM action in the presence of a
stationary symmetry reveals that the
perturbations giving rise to non-vanishing angular
momentum are governed by a self-adjoint system of equations for a
set of gauge invariant fluctuations~\cite{BH97PRD}. With a soliton background,
the solutions to the perturbation equations describe
charged, rotating excitations of the Bartnik--McKinnon
solitons~\cite{BK88soliton}.
In the black-hole case the excitations are combinations
of two branches of stationary perturbations: The first branch comprises
charged black holes with vanishing angular
momentum,%
\epubtkFootnote{As already mentioned in
Section~\ref{subsect-BEM-SP}, these black holes present counter-examples to the
naive generalization of the staticity theorem; they
are nice illustrations of the correct non-Abelian version
of the theorem~\cite{SW92stat, SW93stat}.}
whereas the second one consists of
neutral black holes with non-vanishing angular
momentum. ({A particular combination of the charged and the
rotating branch was found in~\cite{SV97rot}.)}
In the presence of bosonic matter, such as Higgs fields, the
slowly rotating solitons cease to exist, and the two branches
of black-hole excitations merge to a single one with a prescribed
relation between charge and angular momentum~\cite{BH97PRD}. More
information about the EYM--Higgs system can be found
in~\cite{Kunzle:2006ye, OliynykBPS}.

\newpage

\section{Stationary Spacetimes}
\label{sec-SST}

For physical reasons, the black-hole equilibrium states
are expected to be stationary. Spacetimes admitting a Killing
symmetry exhibit a variety of interesting features, some of which
will be discussed in this section. In particular, the existence of a
Killing field implies a canonical local 3+1 decomposition of the
metric. The projection formalism arising from this structure
was developed by Geroch in the early seventies~\cite{Geroch71proj,
  Geroch72proj}, and can be found in Chapter~16 of the book on exact
solutions by Kramer et al.~\cite{Krameretal80exsol}.

A slightly different, rather powerful approach to stationary
spacetimes is obtained by taking advantage of their
Kaluza--Klein (KK) structure. As this approach is less commonly
used in the present context, we will discuss the KK reduction
of the Einstein--Hilbert(--Maxwell) action in some detail,
the more so as this yields an efficient
derivation of the Ernst equations and the Mazur identity.
Moreover, the inclusion of
\textit{non}-Abelian gauge fields within this framework~\cite{BH97PRD}
reveals a decisive structural difference between the
Einstein--Maxwell (EM) and the Einstein--Yang--Mills (EYM)
system. 

\subsection{Reduction of the Einstein--Hilbert action}
\label{subsec-REHA}

By definition, a stationary spacetime $(M,\fourg)$ admits an
asymptotically--time-like Killing field, that is, a vector
field $k$ with $L_{k} \fourg = 0$, $L_{k}$
denoting the Lie derivative with respect to $k$.
At least locally, $M$ has the structure $\Sigma \times G$, where
$G \approx \bbbr$ denotes the one-dimensional group generated by the
Killing symmetry, and $\Sigma$ is the three-dimensional quotient
space $M/G$. A stationary spacetime is called \textit{static}, if
the integral trajectories of $k$ are orthogonal to $\Sigma$.

With respect to an adapted coordinate $t$, so that
$k := \partial_t$, the metric of a stationary
spacetime can be parameterized in terms of a three-dimensional
(Riemannian) metric $\barg := \bar{g}_{ij} \D x^i \D x^j$, a
\newcommand{\MHsigma}{V}
one-form $a := a_i \D x^i$, and a scalar field $\MHsigma$, where
stationarity implies that $\bar{g}_{ij}$, $a_i$ and $\MHsigma$ are
functions on $(\Sigma, \barg)$:
\begin{equation}
\fourg = - \MHsigma (\D t + a)^2 +
\frac{1}{\MHsigma} \, \barg .
\label{REHA-1}
\end{equation}

The notation $t$ suggests that $t$ is a time coordinate,
$\fourgx(\nabla t,\nabla t)<0$, but this restriction does not play any
role in the local form of the equations that we are about to
derive. Similarly the local calculations that follow remain valid
regardless of the causal character of $k$, provided that $ k$ is
\emph{not} null everywhere, and then one only considers the region
where $\fourgx(k,k)\equiv -V$ does not change sign. On any connected
component of this region $k$ is either spacelike or timelike, as
determined by the sign of $V$, and then the metric $\barg$ is
Lorentzian, respectively Riemannian, there. In any case, both the
parameterization of the metric and the equations become singular at
places where $V$ has zeros, so special care is required wherever this
occurs.

Using Cartan's structure equations (see, e.g.,~\cite{NS84LN}),
it is a straightforward task to compute the Ricci scalar for the
above decomposition of the spacetime metric; see, e.g.,~\cite{MH97BHF}
for the details of the derivation. The result is that the
Einstein--Hilbert action of a stationary spacetime reduces to the
action for a scalar field $\MHsigma$ and a vector field $a$, which are
coupled to three-dimensional gravity. The fact that this coupling
is \textit{minimal} is a consequence of the particular choice of the
conformal factor in front of the three-metric
$\barg$ in the decomposition~(\ref{REHA-1}).
The vacuum field equations are thus seen to be
equivalent to the three-dimensional Einstein-matter equations
obtained from variations of the effective action
\begin{equation}
S_{\mathrm{eff}} = \int \barast \left(
\barR - \frac{1}{2 \MHsigma^2} \,
\barsprod{\D \MHsigma}{\D \MHsigma} +
\frac{\MHsigma^2}{2} \,
\barsprod{\D a}{\D a} \right) \, ,
\label{REHA-2}
\end{equation}
with respect to $\bar{g}_{ij}$, $\MHsigma$ and $a$. Here and in the
following $\bar{R}$ denotes the Ricci scalar of $\barg$, while for
$p$-forms $\alpha$ and $\beta$, their inner product is defined by
$\barast \barsprod{\alpha}{\beta} := \alpha \wedge \barast \beta$,
where $\barast$ is the Hodge dual with respect to $\barg$.

It is worth noting that the quantities $\MHsigma$ and $a$
are related to the norm and the twist of the Killing field
as follows:
\begin{equation}
\MHsigma = -\fourg(k,k) \, , \quad
\omega := \frac{1}{2} \fourast( k \wedge \D k)
= - \frac{1}{2} \MHsigma^2 \barast \, \D a \, ,
\label{REHA-3}
\end{equation}
where $\fourast$ and $\barast$ denote the Hodge dual with respect
to $\fourg$ and $\barg$, respectively. {Here and in the
following we use the symbol $k$ for both the Killing field
$\partial_t$ and the corresponding one-form $-\MHsigma(\D t+a)$}.
One can view $a$ as a connection on a principal bundle with base
space $\Sigma$ and fiber $\R$, since it behaves like an Abelian
gauge potential under coordinate transformations of the
form $t \rightarrow t + \varphi(x^i)$. Not surprisingly, it enters
the effective action in a gauge-invariant way, that is,
only via the ``Abelian field strength'', $f := \D a$.

\subsection{The coset structure of vacuum gravity}
\label{subsec-TSVG}

For many applications, in particular for the black-hole
uniqueness theorems, it is convenient to replace
the one-form $a$ by a function, namely the
twist potential. We have already pointed out
that $a$, parameterizing the non-static part of the
metric, enters the effective action~(\ref{REHA-2})
only via the field strength, $f = \D a$. For this reason,
the variational equation for $a$ (that is, the off-diagonal
Einstein equation) takes in vacuum the form of a source-free Maxwell
equation:
\begin{equation}
\D \barast \left(\MHsigma^{2} \/ \D a \right) = 0 \;
\Longrightarrow \;
\D Y = - \barast \left(\MHsigma^{2} \/ \D a \right) \, .
\label{TSVG-1}
\end{equation}
By virtue of Eq.~(\ref{REHA-3}), the (locally-defined) function
$Y$ is a potential for the twist one-form, $\D Y = 2\omega$. In order
to write the effective action~(\ref{REHA-2}) in terms of the twist
potential $Y$, rather than the one-form $a$, one considers
$f $ as a fundamental field and imposes the constraint
$\D f = 0$ with the Lagrange multiplier $Y$. The variational equation
with respect to $f$ then yields $f = -\barast (\MHsigma^{-2} \D Y)$,
which is used to eliminate $f$ in favor of $Y$. One finds
$\frac{1}{2} \MHsigma^2 f \wedge \barast f - Y \D f \rightarrow
-\frac{1}{2} \MHsigma^{-2} \D Y \wedge \barast \D Y$.
Thus, the action~(\ref{REHA-2}) becomes
\begin{equation}
S_{\mathrm{eff}} = \int \barast \left( \barR -
\frac{\barsprod{\D\MHsigma}{\D\MHsigma} +
\barsprod{\D Y}{\D Y}}{2 \MHsigma^2} \right) \, ,
\label{TSVG-2}
\end{equation}
where we recall that $\barsprod{\,}{\,}$ is the inner product with
respect to the three-metric $\barg$ defined in Eq.~(\ref{REHA-1}).

The action~(\ref{TSVG-2}) describes a harmonic map into
a two-dimensional target space, effectively coupled to
three-dimensional gravity. In terms of the complex
Ernst potential $\erpot$~\cite{Ernst68a, Ernst68b},
one has
\begin{equation}
S_{\mathrm{eff}} = \int \barast \left( \barR - 2
\frac{\barsprod{\D\erpot}{\D\bar{\erpot}}}{(\erpot
+ \bar{\erpot})^{2}} \right) \, ,
\quad \erpot := \MHsigma + i \/ Y \, .
\label{TSVG-3}
\end{equation}
The stationary vacuum equations are obtained from variations
with respect to the three-metric $\barg$ [$(ij)$-equations] and the
Ernst potential $\erpot$ [$(0\mu)$-equations]. One easily finds
$\barR_{ij} = 2 (\erpot + \bar{\erpot})^{-2}
\erpot,_{i} \bar{\erpot},_{j}$ and
$\barlap \erpot = 2 (\erpot + \bar{\erpot})^{-1}
\barsprod{\D\erpot}{\D\erpot}$,
where $\barlap$ is the Laplacian with respect to $\barg$.

The target space for stationary vacuum gravity, parameterized by
the Ernst potential $\erpot$, is a K\"ahler manifold with metric
$G_{\erpot \bar{\erpot}} = \partial_{\erpot} \partial_{\bar{\erpot}}
\ln(\MHsigma)$ (see~\cite{DG96square} for details).
By virtue of the mapping
\begin{equation}
\erpot \mapsto z = \frac{1 - \erpot}{1 + \erpot} \, ,
\label{TSVG-4}
\end{equation}
the semi-plane where the Killing field is time-like,
$\mbox{Re}(\erpot)>0$, is mapped into the interior of
the complex unit disc,
$D = \{ z \in \mathbb{C} \mid \, |z| < 1 \}$, with
standard metric
$(1-|z|^2)^{-2} \barsprod{\D z}{\D\bar{z}}$.
By virtue of the stereographic projection,
$\mbox{Re}(z)=x^1(x^0+1)^{-1}$,
$\mbox{Im}(z)=x^2(x^0+1)^{-1}$, the unit disc $D$
is isometric to the pseudo-sphere,
$PS^2 = \{ (x^0,x^1,x^2) \in \bbbr^3 \mid \,
-(x^0)^2 + (x^1)^2 + (x^2)^2 = - 1 \}$.
As the three-dimensional Lorentz group, $SO(2,1)$, acts
transitively and isometrically on the pseudo-sphere with
isotropy group $SO(2)$, the target space is the
coset $PS^2 \approx SO(2,1)/SO(2)$ (see, e.g.,~\cite{KobaNomi69}
or~\cite{B75IDM} for the general theory of symmetric spaces). Using
the universal covering $SU(1,1)$ of $SO(2,1)$, one can parameterize
$PS^2 \approx SU(1,1)/U(1)$ in terms of a positive hermitian matrix
$\Phi(x)$, defined by
\begin{equation}
\Phi(x) = \left(
\begin{array}{cc}
x^0 & x^1 + i\,x^2 \\
x^1 - i\,x^2 & x^0
\end{array}
\right) = \frac{1}{1-|z|^2} \left(
\begin{array}{cc}
1+|z|^2 & 2 \, z \\
2 \bar{z} & 1+|z|^2
\end{array} \right) .
\label{TSVG-5}
\end{equation}
Hence, the effective action for stationary vacuum gravity
becomes the standard action for a $\sigma$-model coupled
to three-dimensional gravity~\cite{NK69sigma},
%
\begin{equation}
{\cal S}_{\mathrm{eff}} = \int \barast \left( \barR -
\frac{1}{4} \, \mbox{Trace} \barsprod{{\cal J}}{{\cal J}}
\right)
 \, ,
\label{act-sigma-stationary}
\end{equation}
where
\begin{equation}
\label{defTrace}
\mbox{Trace} \barsprod{{\cal J}}{{\cal J}} \equiv
\barsprod{{\cal J}^{A}_{\; B}}{{\cal J}^{B}_{\; A}} := \bar{g}^{ij}
({\cal J}_{i})^{A}_{\; B} ({\cal J}_{j})^{B}_{\; A}\,,
\end{equation}
and the currents ${\cal J}_i$ are defined as
\begin{equation}
\label{defCurrent}
{\cal J}_{i} := \Phi^{-1} \bar{\nabla}_{i} \Phi\,.
\end{equation}

The simplest nontrivial solution to the vacuum Einstein equations is
obtained in the static, spherically-symmetric case: For $\erpot =
\MHsigma(r)$ one has $2\bar{R}_{rr} = (\MHsigma'/\MHsigma)^2$ and
$\barlap \ln (\MHsigma) = 0$. With respect to the general
spherically-symmetric ansatz
\begin{equation}
\barg = \D r^2 + \rho^2(r) \D \Omega^2 ,
\label{TSVG-6}
\end{equation}
one immediately obtains the equations
$-4 \rho''/\rho = (\MHsigma'/\MHsigma)^2$ and
$(\rho^2 \MHsigma'/\MHsigma)' = 0$, the solution of
which is the Schwarzschild metric in the usual parametrization:
$\MHsigma = 1 - 2M/r$, $\rho^2 = \MHsigma(r) r^2$.

\subsection{Stationary gauge fields}
\label{subsec-SGF}

The reduction of the Einstein--Hilbert action in the
presence of a Killing field yields a $\sigma$-model,
which is effectively coupled to three-dimensional gravity.
While this structure is retained for the EM system,
it ceases to exist for self-gravitating non-Abelian gauge
fields. In order to perform the dimensional reduction
for the EM and the EYM equations, we need to recall
the notion of a symmetric gauge field.

In mathematical terms, a gauge field (with gauge group $G$, say)
is a connection in a principal bundle $P(M,G)$ over spacetime $M$.
A gauge field is \emph{symmetric} with respect to the action of a
symmetry group $S$ of $M$, if it is described by an $S$-invariant
connection on $P(M,G)$. Hence, finding the symmetric gauge fields
involves the task of classifying the principal bundles $P(M,G)$,
which admit the symmetry group $S$, acting by bundle automorphisms.
This program was carried out by Brodbeck and Straumann
for arbitrary gauge and symmetry groups~\cite{OB95PHD},
(see also~\cite{OB96HPA, BS93Birk}),
generalizing earlier work of Harnad et al. \cite{HSV84},
Jadczyk~\cite{AJ84} and K\"unzle~\cite{HPK94DG}.

The gauge fields constructed in the above way are invariant
under the action of $S$ up to gauge transformations.
This is also the starting point of the alternative approach to the
problem, due to Forg\'acs and Manton~\cite{FM80STS}.
It implies that a gauge potential $A$ is
symmetric with respect to the action of a Killing
field $\xi$, say, if there exists a Lie algebra valued
function ${\cal V}_{\xi}$, such that
\begin{equation}
L_{\xi} A = \rmD \, {\cal V}_{\xi} \, ,
\label{SGF-1}
\end{equation}
where ${\cal V}_{\xi}$ is the generator of an infinitesimal gauge
transformation, $L_{\xi}$ denotes the Lie derivative,
and $\rmD$ is the gauge covariant exterior derivative,
$\rmD {\cal V}_{\xi} = \D {\cal V}_{\xi} + [A,{\cal V}_{\xi}]$.

Let us now consider a stationary spacetime with
(asymptotically) time-like Killing field $k$.
A stationary gauge potential can be parameterized
in terms of a one-form $\barA$ orthogonal to $k$, in the sense that
$\barA (k) = 0$, and a Lie algebra valued potential $\phi$,
\begin{equation}
A = \phi \, (\D t + a) + \barA \,,
\label{SGF-2}
\end{equation}
where we recall that $a$ is the non-static part of the
metric~(\ref{REHA-1}). For the sake of simplicity we adopt a gauge
where ${\cal V}_{k}$ vanishes.%
\epubtkFootnote{The symmetry condition~(\ref{SGF-1}) translates into
  $L_k \phi = [\phi,{\cal V}]$ and $L_k \barA = \barD {\cal V}$, which
  can be used to reduce the EYM equations in the presence of a Killing
  symmetry in a gauge-invariant manner~\cite{MHNS93CQG, MHNS93PLB}.}
By virtue of the above decomposition, the field strength becomes $F =
\barD \phi \wedge (\D t + a) + (\barF + \phi f)$, where $\barF$ is the
Yang--Mills field strength for $\barA$ and $f = da$. Using the
expression~(\ref{TSVG-2}) for the vacuum action, one easily finds that
the EYM action,
\begin{equation}
S_{\mathrm{EYM}} = \int \left( \ast R - 2 \,
\Trace{F \wedge \ast F} \right) \,,
\label{SGF-3}
\end{equation}
where  $R$ and $*$ refer to the 4-dimensional space-time metric $\fourg$ and $\Trace{\,}$ denotes a suitably normalized
trace (e.g.\ $\Trace{\tau_a \tau_b} = \frac{1}{2}\delta_{ab}$
for $SU(2)$, with $\tau_a := \sigma_a/2$,
where the $\sigma_a$'s are the Pauli matrices), gives rise to the effective
action
\begin{equation}
S_{\mathrm{eff}} = \int \barast \left( \barR -
\frac{1}{2 \MHsigma^2} | \D \MHsigma |^{2} +
\frac{\MHsigma^2}{2} | f |^{2} +
\frac{2}{\MHsigma} | \barD \phi |^{2} -
2 \MHsigma | \barF + \phi \, f |^{2} \right) \,,
\label{SGF-4}
\end{equation}
where $\barD$ is the gauge covariant derivative with respect
to $\barA$, and where the inner product also involves the trace:
$\barast | \barF |^{2} := \Trace{\barF \wedge \barast \barF}$.
The above action describes two scalar fields, $\MHsigma$ and $\phi$, and
two vector fields, $a$ and $\barA$, which are minimally coupled to
three-dimensional gravity with metric $\barg$.
Similarly to the vacuum case, the connection $a$ enters $S_{\mathrm{eff}}$
only via the field strength $f $. Again, this gives
rise to a differential conservation law,
\begin{equation}
\D \barast \left[ \MHsigma^2 f - 4 \MHsigma \,
\Trace{ \phi (\barF + \phi f)} \right] = 0 ,
\label{SGF-5}
\end{equation}
by virtue of which one can (locally) introduce a generalized twist
potential $Y$, defined by $-\D Y = \barast [\ldots]$.

The main difference between the Abelian and the non-Abelian
case
concerns the variational equation for $\barA$, that is, the
Yang--Mills equation for $\barF$:
For non-Abelian gauge groups, $\barF$ is no longer an exact
two-form, and the gauge covariant derivative of $\phi$
introduces source terms in the corresponding Yang--Mills
equation:
\begin{equation}
\barD \left[ \MHsigma \barast \left(
\barF + \phi f \right) \right] =
\MHsigma^{-1} \barast \left[ \phi \, , \, \barD \phi \right] .
\label{SGF-6}
\end{equation}
Hence, the scalar magnetic potential -- which can be introduced
in the Abelian case according to
$d \psi := \MHsigma \barast (\barF + \phi f)$ -- ceases
to exist for non-Abelian Yang--Mills fields.
The remaining stationary EYM equations are easily derived from
variations of $S_{\mathrm{eff}}$ with respect to
the gravitational potential $\MHsigma$, the electric Yang--Mills
potential $\phi$ and the three-metric $\barg$.

As an application, we note that the effective action~(\ref{SGF-4})
is particularly suited for analyzing stationary perturbations
of static ($a = 0$), purely magnetic ($\phi = 0$)
configurations~\cite{BH97PRD}, such as the
Bartnik--McKinnon solitons~\cite{BK88soliton}
and the corresponding black-hole solutions~\cite{VG89firsthair,
 KMuA90firsthair, PB90firsthair}. The two crucial observations in
this context are~\cite{BH97PRD, SV97rot}:

\begin{enumerate}[(i)]
\item The only perturbations of the static, purely magnetic
EYM solutions, which can contribute the ADM
angular momentum are the purely non-static,
purely electric ones, $\delta a$ and $\delta \phi$.

\item In first-order perturbation theory, the relevant
fluctuations, $\delta a$ and $\delta \phi$,
decouple from the remaining metric and matter perturbations.
\end{enumerate}

The second observation follows from the fact that
the magnetic Yang--Mills equation~(\ref{SGF-6}) and the
Einstein equations for $\MHsigma$ and $\barg$
become background equations, since they contain no linear
terms in $\delta a$ and $\delta \phi$. Therefore, the purely electric,
non-static perturbations are
governed by the twist equation~(\ref{SGF-5})
and the electric Yang--Mills equation (obtained
from variations of $S_{\mathrm{eff}}$ with respect to $\phi$).

Using Eq.~(\ref{SGF-5}) to introduce the twist potential $Y$,
the fluctuation equations for the first-order quantities
$\delta Y$ and $\delta \phi$ assume the form of a
self-adjoint system~\cite{BH97PRD}.
Considering perturbations of spherically-symmetric configurations, one
can expand $\delta Y$ and $\delta \phi$ in terms of isospin
harmonics. In this way one obtains a Sturm--Liouville problem, the
solutions of which reveal the features mentioned in the last paragraph
of Section~\ref{subsect-BEM-RBH}~\cite{BHSV97}.

\subsection{The stationary Einstein--Maxwell system}
\label{subsec-SEMS}

In the one-dimensional Abelian case, both the off-diagonal
Einstein equation~(\ref{SGF-5}) \textit{and} the Maxwell
equation~(\ref{SGF-6}) give rise to scalar potentials,
(locally) defined by
\begin{equation}
\D \psi := \MHsigma \barast \left(\barF + \phi f \right) \,,
\quad
\D Y := -\MHsigma^{2} \barast f +
2\phi \/ \D \psi - 2\psi \/ \D \phi \,.
\label{SEMS-1}
\end{equation}
Similarly to the vacuum case, this enables one to apply the Lagrange
multiplier method to express the effective action in terms of the
scalar fields $Y$ and $\psi$, rather than the one-forms $a$ and
$\barA$. It turns out that in the stationary-axisymmetric case, to
which we return in Section~\ref{sect-SAST}, we will also be interested
in the dimensional reduction of the EM system with respect to a
\textit{space-like} Killing field. Therefore, we give here the general
result for an arbitrary Killing field $\xi$ with norm $N$:
\begin{equation}
S_{\mathrm{eff}} = \int \barast \left( \barR - 2
\frac{| \D \phi |^{2} + | \D \psi |^{2}}{N} -
\frac{| \D N |^{2} + | \D Y - 2 \phi \D \psi +
2 \psi \D \phi |^{2}}{2 \; N^{2}} \right) \,,
\label{SEMS-2}
\end{equation}
where $\barast |\D \phi|^2 =\D \phi \wedge \barast \D \phi$, etc.
The electro-magnetic potentials $\phi$ and $\psi$ and the
gravitational scalars $N$ and $Y$ are obtained from the
four-dimensional field strength $F$ and the Killing field
as follows (given a two-form
$\beta$, we denote by $i_{\xi} \beta$ the one-form with components
$\xi^{\mu} \beta_{\mu \nu}$):
\begin{equation}
\D \phi = - i_{\xi} \fourF \, , \quad
\D \psi = i_{\xi} \fourast \fourF \,,
\label{SEMS-3}
\end{equation}
\begin{equation}
N = \sprod{\xi}{\xi} \, , \quad
\D Y = 2 \left( \omega + \phi \/ \D \psi - \psi \/ \D \phi \right) \,,
\label{SEMS-4}
\end{equation}
where $2 \omega := \fourast (\xi \wedge \D \xi)$. The inner product
$\barsprod{\cdot}{\cdot}$ and the associated ``norm'' $|\cdot|$ are
taken with respect to the three-metric $\barg$, which becomes
pseudo-Riemannian if $\xi$ is space-like. The additional stationary
symmetry will then imply that the inner products in~(\ref{SEMS-2})
have a fixed sign, despite the fact that $\barg$ is not a Riemannian
metric in this case.

The action~(\ref{SEMS-2}) describes a harmonic mapping into a
four-dimensional target space, effectively coupled to
three-dimensional gravity. In terms of
the complex Ernst potentials, $\Lambda := - \phi + i \psi$ and
$\erpot := - N - \Lambda \bar{\Lambda} + i Y$~\cite{Ernst68a,
  Ernst68b}, the effective EM action becomes
\begin{equation}
S_{\mathrm{eff}} = \int \barast \left( \barR - 2
\frac{\mid \D \Lambda \mid^{2}}{N} -
\frac{1}{2} \,
\frac{\mid \D \erpot + 2 \bar{\Lambda} \D \Lambda \mid^{2}}{N^{2}}
\right) \,,
\label{SEMS-5}
\end{equation}
where $| \D \Lambda |^{2} :=
\barsprod{\D \Lambda}{\overline{\D \Lambda}}$.
The field equations are obtained from variations
with respect to the three-metric $\barg$ and the
Ernst potentials. In particular, the equations for
$\erpot$ and $\Lambda$ become
\begin{equation}
\barlap \erpot = -
\frac{\barsprod{\D \erpot}{\D \erpot + 2
\bar{\Lambda} \D \Lambda}}{N} \,, \quad
\barlap \Lambda = -
\frac{\barsprod{\D \Lambda}{\D \erpot + 2
\bar{\Lambda} \D \Lambda}}{N} \,,
\label{SEMS-6}
\end{equation}
where $-N = \Lambda \bar{\Lambda} +
\frac{1}{2}(\erpot+\bar{\erpot})$. The isometries of the target
manifold are obtained by solving the respective Killing
equations~\cite{NK69sigma} (see also~\cite{Kin77, KinC77, KinC78a,
  KinC78b}). This reveals the coset structure of the target space and
provides a parametrization of the latter in terms of the Ernst
potentials. For vacuum gravity and a timelike Killing vector we have
seen in Section~\ref{subsec-TSVG} that the coset space, $G/H$, is
$SU(1,1)/U(1)$, whereas one finds $G/H = SU(2,1)/S(U(1,1) \times
U(1))$ for the stationary EM equations. If the dimensional reduction
is performed with respect to a space-like Killing field, then $G/H =
SU(2,1)/ S(U(2)\times U(1))$. The explicit representation of the coset
manifold in terms of the above Ernst potentials, $\erpot$ and
$\Lambda$, is given by the Hermitian matrix $\Phi$, with components
\begin{equation}
\Phi_{A B} = \eta_{A B} + 2 \mbox{sig}(N) \bar{v}_A v_B \,,
\quad
(v_0 \/ , v_1 \/ , v_2) :=
\frac{1}{2 \sqrt{| N |}}
( \erpot-1 , \erpot+1 , 2 \Lambda ) \,,
\label{SEMS-7}
\end{equation}
where $v_{A}$ is the Kinnersley
vector~\cite{Kin73},
and $\eta := \mbox{diag} (-1,+1,+1)$.
It is straightforward to verify that, in terms of $\Phi$,
the effective action~(\ref{SEMS-5}) assumes the
$SU(2,1)$ invariant form~\eqref{act-sigma-stationary}.
The equations of motion following from this action are
the following three-dimensional Einstein equations with sources, obtained from
variations with respect to $\barg$, and the $\sigma$-model
equations, obtained from variations with respect to $\Phi$:
\begin{equation}
\barR_{ij} = \frac{1}{4} \,
\trMH{ {\cal J}_{i} \, {\cal J}_{j} } \, ,
\quad \D \barast {\cal J} = 0 \,;
\label{SEMS-9}
\end{equation}
here all operations are taken with respect to $\barg$.

\newpage

\section{Some Applications}
\label{sect-ACS}

The $\sigma$-model structure is responsible for various distinguished
features of the stationary EM system and related
self-gravitating matter models.
This section is devoted to a brief discussion of some
applications. We show how the Mazur identity~\cite{PM84b},
the quadratic mass formulae~\cite{MH97MF} and the
Israel--Wilson--Perj\'es  class of stationary black holes~\cite{IW72Class, ZP71}
arise from the $\sigma$-model structure of the stationary
field equations.

\subsection{The Mazur identity}
\label{subsec-TMI}

In the presence of a second Killing field, the EM
equations~(\ref{SEMS-9}) experience further, considerable
simplifications, which will be discussed later. In this section we
will not yet require the existence of an additional Killing
symmetry. The Mazur identity~\cite{PM84b}, which is the key to the
uniqueness theorem for the Kerr--Newman metric~\cite{PM82UT, PM84a},
is a consequence of the coset structure of the field equations. Note,
however, that while the derivation of the general form of this
identity only requires one Killing vector, its application to the
uniqueness argument uses two; we will return to this issue shortly.

In order to obtain the Mazur identity, one considers two arbitrary
Hermitian matrices, $\Phi_{1}$ and $\Phi_{2}$. The aim is to compute
the Laplacian with respect to a metric $\barg$ (which in the
application of interest will be flat) of the relative difference
$\Psi$, say, between $\Phi_{2}$ and $\Phi_{1}$,
\begin{equation}
\Psi := \Phi_{2} \Phi_{1}^{-1} - \mathbb{1}\,.
\label{TMI-1}
\end{equation}
It turns out to be convenient to introduce the current matrices
${\cal J}_{1} = \Phi^{-1}_{1} \bar{\nabla} \Phi_{1}$ and
${\cal J}_{2} = \Phi^{-1}_{2} \bar{\nabla} \Phi_{2}$,
and their difference ${\cal J}_{\triangle} = {\cal J}_{2} - {\cal J}_{1}$,
where $\bar{\nabla}$ denotes the covariant derivative with respect
to $\barg$. Using
$\bar{\nabla} \Psi = \Phi_{2} \, {\cal J}_{\triangle} \, \Phi_{1}^{-1}$,
the Laplacian of $\Psi$ becomes
\begin{displaymath}
\bar{\Delta} \Psi =
\barsprod{\bar{\nabla} \Phi_{2}}{{\cal J}_{\triangle}} \, \Phi_{1}^{-1} +
\Phi_{2} \, \barsprod{{\cal J}_{\triangle}}{\bar{\nabla} \Phi_{1}^{-1}} +
\Phi_{2} \, (\bar{\nabla} {\cal J}_{\triangle}) \, \Phi_{1}^{-1} \,,
\end{displaymath}
where, as before, the inner product $\barsprod{\cdot}{\cdot}$ is
taken with respect to the three-metric $\barg$ and also involves a
matrix multiplication. For hermitian matrices one has
$\bar{\nabla} \Phi_{2} = {\cal J}_{2}^{\dagger}\Phi_{2}$ and
$\bar{\nabla} \Phi_{1}^{-1} = -\Phi_{1}^{-1} {\cal J}_{1}^{\dagger}$, which
can be used to combine the trace of the first two terms
on the right-hand side of the above expression. One easily finds
\begin{equation}
\trMH{ \bar{\Delta} \Psi} =
\trMH{ \barsprod{\Phi_{1}^{-1}
{\cal J}_{\triangle}^{\dagger}}{\Phi_{2} {\cal J}_{\triangle}} +
\Phi_{2} \, (\bar{\nabla} {\cal J}_{\triangle}) \, \Phi_{1}^{-1}} \,.
\label{TMI-2}
\end{equation}
The above expression is an identity for the relative difference of two
arbitrary Hermitian matrices, with all operations taken with respect
to $\barg$ (recall~\eqref{defTrace}). If the latter are
\textit{solutions} of a non-linear $\sigma$-model with action $\int
\trMH{{\cal J} \wedge \barast {\cal J}}$, then their currents are
conserved [see Eq.~(\ref{SEMS-9})], implying that the second term on
the right-hand side vanishes. Moreover, if the $\sigma$-model
describes a mapping with coset space $SU(p,q)/S(U(p) \times U(q))$,
then this is parameterized by positive Hermitian matrices of the form
$\Phi = g g^{\dagger}$. (We refer to~\cite{KobaNomi69, EF80BK},
and~\cite{B75IDM} for the theory of symmetric spaces.) Hence, the
``on-shell'' restriction of the Mazur identity to $\sigma$-models with
coset $SU(p,q)/S(U(p) \times U(q))$ becomes
\begin{equation}
\trMH{\bar{\Delta} \Psi} =
\mbox{Trace} \barsprod{{\cal M}}{{\cal M}^{\dagger}} ,
\label{TMI-3}
\end{equation}
where ${\cal M} := g_{1}^{-1} {\cal J}^{\dagger}_{\triangle} g_{2}$.

Of decisive importance to the uniqueness proof for the
Kerr--Newman metric is the fact that the right-hand side of the above
relation is non-negative. In order to achieve this one needs
\textit{two} Killing fields: The requirement that $\Phi$
be represented in the form $g g^{\dagger}$ forces
the reduction of the EM system with respect to a \textit{space-like}
Killing field; otherwise the coset is
$SU(2,1)/S(U(1,1) \times U(1))$, which is not of the desired form.
As a consequence of the space-like reduction,
the three-metric $\barg$ is
not Riemannian, and the right-hand side of Eq.~(\ref{TMI-3}) is
indefinite, unless the matrix valued one-form ${\cal M}$
is space-like. This is the case if there exists a
time-like Killing field with $L_{k} \Phi = 0$,
implying that the currents are orthogonal to $k$:
${\cal J}(k) = i_{k} \Phi^{-1} d \Phi = \Phi^{-1} L_{k} \Phi = 0$.
The reduction of Eq.~(\ref{TMI-3}) with respect to the second Killing
field and the integration of the resulting expression will be
discussed in Section~\ref{sect-SAST}.

\subsection{Mass formulae}
\label{subsec-MF}

The stationary vacuum Einstein equations describe a two-dimensional
$\sigma$-model coupled to three-dimensional
gravity. The target manifold is the pseudo-sphere
$SO(2,1)/SO(2) \approx SU(1,1)/U(1)$, which is
parameterized in terms of the norm and the twist potential of the
Killing field (see Section~\ref{subsec-TSVG}).
The symmetric structure of the target space persists for the
stationary EM system, where the four-dimensional coset,
$SU(2,1)/S(U(1,1) \times U(1))$, is represented by a hermitian
matrix $\Phi$, comprising the two electro-magnetic scalars,
the norm of the Killing field and the generalized twist potential
(see Section~\ref{subsec-SEMS}).

The coset structure of the stationary field equations is shared by
various self-gravitating matter models with massless scalars (moduli)
and Abelian vector fields. For scalar mappings into a symmetric target
space $\bar{G}/\bar{H}$, say, Breitenlohner et al.~\cite{BMG88KK} have
classified the models admitting a symmetry group, which is sufficiently
large to comprise \textit{all} scalar fields arising on the effective
level%
\epubtkFootnote{In addition to the actual scalar fields, the effective
  action comprises two gravitational scalars (the norm and the
  generalized twist potential) and two scalars for each stationary
  Abelian vector field (electric and magnetic potentials).}
within \textit{one} coset space, $G/H$. A prominent example of this
kind is the EM-dilaton-axion system, which is relevant to $N = 4$
supergravity and to the bosonic sector of four-dimensional heterotic
string theory: The pure dilaton-axion system has an $SL(2,\bbbr)$
symmetry, which persists in dilaton-axion gravity with an Abelian gauge
field~\cite{DG96GKC}. Like the EM system, the model also possesses an
$SO(1,2)$ symmetry, arising from the dimensional reduction with
respect to the Abelian isometry group generated by the Killing
field. However, Gal'tsov and Kechkin~\cite{GK94EH, GK95MD} have shown that the
full symmetry group is larger than $SL(2,\bbbr) \times
SO(1,2)$: The target space for dilaton-axion gravity with a $U(1)$
vector field is the coset $SO(2,3)/(SO(2) \times
SO(1,2))$~\cite{DG95IS}. Using the fact that $SO(2,3)$ is isomorphic
to $Sp(4,\bbbr)$, Gal'tsov and Kechkin~\cite{GK95U} were also able to
give a parametrization of the target space in terms of $4 \times 4$
(rather than $5 \times 5$) matrices. The relevant coset space was
shown to be $Sp(4,\bbbr)/U(1,1)$; for the generalization to the
dilaton-axion system with multiple vector fields we refer
to~\cite{GL96EH, GS97}.

Common to the black-hole solutions of the above models is the fact
that their Komar mass can be expressed in terms of the total charges
and the area and surface gravity of the horizon~\cite{MH97MF}. The
reason for this is the following: Like the EM
equations~(\ref{SEMS-9}), the stationary field equations consist of
the three-dimensional Einstein equations and the $\sigma$-model
equations,
\begin{equation}
\barR_{ij} = \frac{1}{4} \,
\trMH{{\cal J}_{i} \, {\cal J}_{j}} \,,
\quad \D \barast {\cal J} = 0 \,.
\label{MF-1}
\end{equation}
The current one-form ${\cal J} := \Phi^{-1} \D \Phi$
is given in terms of the Hermitian matrix $\Phi$, which
comprises all scalar fields arising on the effective level.
The $\sigma$-model equations, $\D \bar{\ast} {\cal J} = 0$,
include $\mbox{dim}(G)$ differential current conservation laws, of
which $\mbox{dim}(H)$ are redundant. Integrating all equations over
a space-like hypersurface extending from the horizon to infinity,
Stokes' theorem yields a set of relations between the charges and the
horizon-values of the scalar potentials. A very familiar
relation of this kind is the Smarr formula~\cite{LS73MF};
see Eq.~(\ref{MF-5}) below.
The crucial observation is that Stokes' theorem provides
$\mbox{dim}(G)$ \textit{independent} Smarr relations,
rather than only $\mbox{dim}(G/H)$ ones.
(This is due to the fact that all $\sigma$-model currents are
\textit{algebraically} independent, although there are
$\mbox{dim}(H)$ differential identities,
which can be derived from the $\mbox{dim}(G/H)$ field
equations.)

The \textit{complete} set of Smarr type formulae can be used to get
rid of the horizon-values of the scalar potentials. In this way one
obtains a relation, which involves only the Komar mass, the charges and
the horizon quantities. For the EM-dilaton-axion system one finds, for
instance~\cite{MH97MF},
\begin{equation}
\left(\frac{1}{4 \pi} \/ \kappa \/ {\cal A} \right)^{2} =
M^{2} + N^{2} + D^{2} + A^{2} - Q^{2} - P^{2} ,
\label{MF-2}
\end{equation}
where $\kappa$ and ${\cal A}$ are the surface gravity and the area of
the horizon, and the right-hand side comprises the asymptotic flux
integrals, that is, the total mass, the NUT charge, the dilaton and
axion charges, and the electric and magnetic charges,
respectively. The derivation of Eq.~(\ref{MF-2}) is not restricted to
static configurations. However, when evaluating the surface terms, one
assumes that the horizon is generated by the same Killing field that
is also used in the dimensional reduction; the asymptotically
time-like Killing field $k$. A generalization of the method to
rotating black holes requires the evaluation of the potentials
(defined with respect to $k$) on a Killing horizon, which is generated
by $\ell = k + \Omega_H m$, rather than $k$.

A very simple illustration of the idea outlined above is the static,
purely electric EM system. In this case, the electrovacuum coset
$SU(2,1)/S(U(1,1) \times U(1))$ reduces to $G/H = SU(1,1)/\bbbr$. The
matrix $\Phi$ is parameterized in terms of the electric potential
$\phi$ and the gravitational potential $\MHsigma := - k_{\mu}
k^{\mu}$. The $\sigma$-model equations comprise $\mbox{dim}(G)=3$
differential conservation laws, of which $\mbox{dim}(H)=1$ is
redundant:
\begin{equation}
\D \bar{\ast} \left( \frac{\D \phi}{\MHsigma} \right) = 0 , \quad
\D \bar{\ast} \left( \frac{\D \MHsigma}{\MHsigma} - 2 \phi \,
\frac{\D \phi}{\MHsigma} \right) = 0 ,
\label{MF-3}
\end{equation}
\begin{equation}
\D \bar{\ast} \left( \left( \MHsigma + \phi^2 \right) \,
\frac{\D \phi}{\MHsigma} - \phi \,
\frac{\D \MHsigma}{\MHsigma} \right) = 0 .
\label{MF-4}
\end{equation}
[It is immediately verified that Eq.~(\ref{MF-4}) is indeed a
  consequence of the Maxwell and Einstein Eqs.~(\ref{MF-3}).]
Integrating Eqs.~(\ref{MF-3}) over a space-like hypersurface and
using Stokes' theorem yields
\begin{equation}
Q = Q_H \, , \quad
M = \frac{\kappa}{4\pi} {\cal A} + \phi_H Q_H ,
\label{MF-5}
\end{equation}
which is the well-known Smarr formula; to establish it, one also uses
the fact that the electric potential assumes a constant value $\phi_H$
on the horizon. Also, the quantity $Q_H$ is defined by the flux
integral of $\ast F$ over the horizon (at time $\Sigma$), while the
corresponding integral of $\ast \D k$ gives $\kappa {\cal A}/4\pi$
(see~\cite{MH97MF} for details). In a similar way, Eq.~(\ref{MF-4})
provides an \textit{additional} relation of the Smarr type,
\begin{equation}
Q = 2 \phi_H \frac{\kappa}{4\pi} {\cal A} + \phi_H^2 Q_H ,
\label{MF-6}
\end{equation}
which can be used to compute the horizon-value of the electric
potential, $\phi_H$. Using this in the Smarr formula~(\ref{MF-5})
gives the desired expression for the total mass, $M^2 = (\kappa{\cal
  A}/4\pi)^{2}+Q^2$.

In the ``extreme'' case, the Bogomol'nyi--Prasad--Sommerfield (BPS)
bound~\cite{GH82BOG} for the static EM-dilaton-axion system, $0 =
M^{2} + D^{2} + A^{2} - Q^{2} - P^{2}$, was previously obtained by
constructing null geodesics of the target space~\cite{CG96}. For
spherically-symmetric configurations with non-degenerate horizons
($\kappa \neq 0$), Eq.~(\ref{MF-2}) was derived by Breitenlohner et
al.~\cite{BMG88KK}. In fact, many of the spherically-symmetric
black-hole solutions with scalar and vector fields~\cite{GWG82dil,
  GWGM88, GHS91} are known to fulfill Eq.~(\ref{MF-2}), where the
left-hand side is expressed in terms of the horizon radius
(see~\cite{GL97} and references therein). Using the generalized first
law of black-hole thermodynamics, Gibbons et al.~\cite{GKK96} obtained
Eq.~(\ref{MF-2}) for spherically-symmetric solutions with an arbitrary
number of vector and moduli fields.

The above derivation of the mass formula~(\ref{MF-2}) is neither
restricted to spherically-symmetric configurations, nor are the
solutions required to be static. The crucial observation is that the
coset structure gives rise to a set of Smarr formulae, which is
sufficiently large to derive the desired relation. Although the
result~(\ref{MF-2}) was established by using the explicit
representations of the EM and EM-dilaton-axion coset
spaces~\cite{MH97MF}, similar relations are expected to exist in the
general case. More precisely, it should be possible to show that the
Hawking temperature of all asymptotically-flat (or asymptotically NUT)
non-rotating black holes with massless scalars and Abelian vector
fields is given by
\begin{equation}
T_H = \frac{2}{{\cal A}} \, \sqrt{\sum (Q_S)^2 -
\sum (Q_V)^2} \, ,
\label{HTEMP}
\end{equation}
provided that the stationary field equations assume the
form~(\ref{MF-1}), where $\Phi$ is a map into a \textit{symmetric
  space}, $G/H$. Here $Q_S$ and $Q_V$ denote the charges of the
scalars (including the gravitational ones) and the vector fields,
respectively.

\subsection{The Israel--Wilson--Perj\'es class}
\label{subsec-IWC}

A particular class of solutions to the stationary EM equations
is obtained by requiring that the
Riemannian manifold $(\Sigma, \barg)$ is flat~\cite{IW72Class}; see also~\cite{ZP71}.
For $\bar{g}_{ij} = \delta_{ij}$, the three-dimensional Einstein
equations obtained from variations of the effective
action~(\ref{SEMS-5}) with respect to $\barg$ become
\begin{equation}
4 \MHsigma \, \Lambda,_{i} \bar{\Lambda},_{j} =
\left(\erpot,_{i} + 2 \bar{\Lambda} \Lambda,_{i} \right)
\left(\bar{\erpot},_{j} + 2 \Lambda \bar{\Lambda},_{j} \right)\,,
\label{IWC-1}
\end{equation}
where, as we are considering stationary configurations, we use the
dimensional reduction with respect to the asymptotically--time-like
Killing field $k$ with norm $\MHsigma = -\sprod{k}{k} = -N$. Israel
and Wilson~\cite{IW72Class} have shown that all solutions of this
equation fulfill $\Lambda = c_{0} + c_{1} \erpot$. In fact, it is not
hard to verify that this ansatz solves Eq.~(\ref{IWC-1}), provided
that the complex constants $c_{0}$ and $c_{1}$ are subject to $c_{0}
\bar{c}_{1} + c_{1} \bar{c}_{0} = -1/2$. Using asymptotic flatness,
and adopting a gauge where the limits at infinity of the
electro-magnetic potentials and the twist potential vanish, one has
$\erpot_{\infty}:=\lim_{r\rightarrow \infty}\erpot = 1$ and
$\Lambda_{\infty}:=\lim_{r\rightarrow \infty}\Lambda = 0$, and thus
\begin{equation}
\Lambda = \frac{\mbox{e}^{i \alpha}}{2}
\left( 1 - \erpot \right) , \quad
\mbox{where} \; \alpha \in \bbbr.
\label{IW-2}
\end{equation}
It is crucial that this ansatz solves both the equation for $\erpot$
and the one for $\Lambda$: One easily verifies that
Eqs.~(\ref{SEMS-6}) reduce to the single equation
\begin{equation}
\barlap \left( 1+\erpot \right)^{-1} = 0 ,
\label{IW-3}
\end{equation}
where $\barlap$ is the three-dimensional \textit{flat} Laplacian.

For \textit{static, purely electric} configurations
the twist potential $Y$ and the magnetic potential
$\psi$ vanish. The ansatz~(\ref{IW-2}),
together with the definitions of the Ernst potentials,
$\erpot = \MHsigma - |\Lambda|^{2} + iY$
and $\Lambda = -\phi + i \psi$ (see Section~\ref{subsec-SEMS}),
yields
\begin{equation}
1 + \erpot = 2\sqrt{\MHsigma} , \quad
\mbox{and} \quad \phi = 1 -\sqrt{\MHsigma} .
\label{IW-4}
\end{equation}
Since $\MHsigma_{\infty} = 1$, the linear relation between
$\phi$ and the gravitational potential $\sqrt{\MHsigma}$
implies $(\D \MHsigma)_{\infty} = -(2 \D \phi)_{\infty}$.
By virtue of this, the total mass and the total charge of every
asymptotically flat, static, purely electric Israel--Wilson--Perj\'es 
solution are equal:
\begin{equation}
M = -\frac{1}{8\pi}\int\ast dk = -\frac{1}{4\pi}\int\ast F = Q ,
\label{IW-5}
\end{equation}
where the integral extends over an asymptotic
two-sphere. Note that for purely electric configurations
one has $F = k \wedge \D \phi/\MHsigma$; also, staticity implies
$k = -\MHsigma \D t$ and thus $dk = -k \wedge \D \MHsigma/\MHsigma=-F$.
The simplest nontrivial solution of the flat Poisson
equation~(\ref{IW-3}), $\barlap \MHsigma^{-1/2} = 0$,
corresponds to a linear combination of $n$ monopole sources $m_a$
located at arbitrary points $\underline{x}_a$,
\begin{equation}
\MHsigma^{-1/2}(\underline{x}) = 1 + \sum_{a=1}^n \frac{m_a}{|\underline{x} - \underline{x}_a|} .
\label{IW-6}
\end{equation}
This is the MP solution~\cite{P45PM, M47PM},
with spacetime metric $\fourg = -\MHsigma \D t^2 + \MHsigma^{-1} \D
\underline{x}^{2}$ and electric potential
$\phi=1-\sqrt{\MHsigma}$. The MP metric describes a \textit{regular}
black-hole spacetime, where the horizon comprises $n$ disconnected
components. Hartle and Hawking~\cite{SHJH72} have shown that all
singularities are ``hidden'' behind these null surfaces. In Newtonian
terms, the configuration corresponds to $n$ arbitrarily-located
charged mass points with $|q_a| = \sqrt{G}m_a$.  

Non-static members of the Israel--Wilson--Perj\'es class were constructed as
well~\cite{IW72Class, ZP71}. However, these generalizations of the
MP multi--black-hole solutions share certain
unpleasant properties with NUT spacetime~\cite{NUT63} (see
also~\cite{DB64, CM65}). In fact, the results of~\cite{CRT}
(see~\cite{SHJH72, CN95PM, MH97PM} for previous results) suggest that
-- except the MP solutions -- all configurations obtained by the
Israel--Wilson--Perj\'es  technique either fail to be asymptotically flat or have
naked singularities.

\newpage

\section{Stationary and Axisymmetric Spacetimes}
\label{sect-SAST}

The presence of two Killing symmetries yields a considerable
simplification of the field equations. In fact, for certain matter
models the latter become completely integrable~\cite{DM79INT},
provided that the Killing fields satisfy the orthogonal-integrability
conditions. Spacetimes admitting two Killing fields provide the
framework for both the theory of colliding gravitational waves and the
theory of rotating black holes~\cite{SC91BK}. Although dealing with
different physical subjects, the theories are mathematically closely
related. We refer the reader to Chandrasekhar's comparison between
corresponding solutions of the Ernst equations~\cite{SC89Gibbs}.

This section reviews the structure of the stationary and axisymmetric
field equations. We start by recalling the circularity problem. It is
argued that circularity is \textit{not} a generic property of
asymptotically-flat, stationary and axisymmetric spacetimes. However, if the symmetry conditions for the matter fields do imply
circularity, then the reduction with respect to the second Killing
field simplifies the field equations drastically. The systematic
derivation of the Kerr--Newman metric and the proof of its uniqueness
provide impressive illustrations of this fact.

\subsection{Integrability properties of Killing fields}
\label{subsec-CIR}

Our aim here is to discuss the circularity problem in some
more detail. The task is to use the
symmetry properties of the matter model in order to
establish the orthogonal-integrability conditions for the
Killing fields. The link between the relevant components of the
stress-energy tensor and the integrability conditions
is provided by a general identity for the derivative
of the twist of a Killing field $\xi$, say,
\begin{equation}
\D \omega_{\xi} = \ast \left[ \xi \wedge R(\xi) \right],
\label{CIR-1}
\end{equation}
and Einstein's equations, implying
$\xi \wedge R(\xi) = 8 \pi [\xi \wedge T(\xi)]$. This
follows from the definition of the twist and
the Ricci identity for Killing fields,
$\Delta \xi = -2 R(\xi)$, where $R(\xi)$ is the one-form with
components $[R(\xi)]_{\mu} := R_{\mu \nu} \xi^{\nu}$;
see, e.g.,~\cite{MH96LN}, Chapter~2. For a stationary
and axisymmetric spacetime with Killing
fields (one-forms) $k$ and $m$, Eq.~(\ref{CIR-1})
implies
\begin{equation}
\D \/ \sprod{m}{\omega_{k}} = - 8 \pi \fourast
\left[ m \wedge k \wedge T(k) \right] ,
\label{CIR-2}
\end{equation}
and similarly for $k \leftrightarrow m$. Eq.~(\ref{CIR-2}) is an
identity up to a term involving the Lie derivative of the twist of the
first Killing field with respect to the second one (since $\D
\sprod{m}{\omega_{k}} = L_{m}\omega_{k} - i_{m} \D \omega_{k}$). In
order to establish $L_m \omega_k = 0$, it is sufficient to show that
$k$ and $m$ commute in an asymptotically-flat spacetime. This was
first achieved by Carter~\cite{BC70KF} and later, under more general
conditions, by Szabados~\cite{LS87KF}.

The following is understood to also apply for $k \leftrightarrow m$:
By virtue of Eq.~(\ref{CIR-2}) -- and the fact that the condition $m
\wedge k \wedge dk = 0$ can be written as $\sprod{m}{\omega_{k}}=0$ --
the circularity problem is reduced to the following two tasks:

\begin{enumerate}[(i)]
\item Show that $\D \sprod{m}{\omega_{k}}=0$ implies
$\sprod{m}{\omega_{k}}=0$.
\item Establish $m \wedge k \wedge T(k) = 0$ from the stationary and
axisymmetric matter equations.
\end{enumerate}

(i) Since $\sprod{m}{\omega_k}$ is a \textit{function}, it is locally
constant if its derivative vanishes. As $m$ vanishes on the rotation
axis, this implies $\sprod{m}{\omega_k}=0$ in every connected domain
of spacetime intersecting the axis. (At this point it is worthwhile to
recall that the corresponding step in the staticity theorem requires
more effort: Concluding from $\D \omega_k = 0$ that $\omega_k$
vanishes is more involved, since $\omega_k$ is a
\textit{one-form}. However, using the Stokes' theorem to integrate an
identity for the twist~\cite{MH96HPA} shows that a strictly stationary
-- not necessarily simply connected -- domain of outer communication
must be static if $\omega_k$ is closed. While this proves the
staticity theorem for vacuum and self-gravitating scalar
fields~\cite{MH96HPA}, it does not solve the electrovacuum case. It
should be noted that in the context of the proof of uniqueness the
strictly stationary property follows from staticity~\cite{ChGstatic}
and not the other way around (compare
Figure~\ref{fig:classification}).

(ii) While $m \wedge k \wedge T(k) = 0$ follows from the symmetry
conditions for electro-magnetic fields~\cite{BC69KH} and for scalar
fields~\cite{MH95UROT}, it cannot be established for non-Abelian gauge
fields~\cite{MH96HPA}. This implies that the usual foliation of
spacetime used to integrate the stationary and axisymmetric Maxwell
equations is too restrictive to treat the EYM system. This is seen as
follows: In Section~\ref{subsec-SGF} we have derived the
formula~(\ref{SGF-5}). By virtue of Eq.~(\ref{REHA-3}) this becomes an
expression for the derivative of the twist in terms of the electric
Yang--Mills potential $\phi_k$ (defined with respect to the stationary
Killing field $k$) and the magnetic one-form $i_k \ast F = \MHsigma
\barast (\barF + \phi_k f)$:
\begin{equation}
\D \left[ \omega_k + 4 \, \Trace{\phi_k \, i_k \ast F} \right] = 0\,,
\label{CIR-3}
\end{equation}
where $\Trace{\,}$ is a suitably normalized trace (see equation~\eqref{SGF-3}).  Contracting this relation with the axial Killing field $m$, and using
again the fact that the Lie derivative of $\omega_k$ with respect to
$m$ vanishes, yields immediately
\begin{equation}
\D \sprod{m}{\omega_k} = 0 \; \Longleftrightarrow \;
\Trace{\phi_k \, \left(\ast F \right)(k,m)} = 0 \, .
\label{CIR-4}
\end{equation}
The difference between the Abelian and the non-Abelian case is due to
the fact that the Maxwell equations automatically imply that the
$(km)$-component of $\ast F$ vanishes, whereas this does not follow
from the Yang--Mills equations. In fact, the Maxwell equation $\D \ast
F = 0$ and the symmetry property $L_k \ast F = \ast L_k F = 0$ imply
the existence of a magnetic potential, $\D \psi = (\ast F)(k,\, \cdot
\, )$, thus, $(\ast F)(k,m) = i_m \D \psi = L_m \psi = 0$. Moreover,
the latter do not imply that the Lie algebra valued scalars $\phi_k$
and $\left(\ast F \right)(k,m)$ are orthogonal. Hence, circularity is
an intrinsic property of the EM system, whereas it imposes additional
requirements on non-Abelian gauge fields.

Both staticity and circularity theorems can be established for scalar
fields or, more generally, scalar mappings with arbitrary target
manifolds: Consider, for instance, a self-gravitating scalar mapping
$\phi: (M,\fourg) \rightarrow (N,\targetg)$ with Lagrangian $L[\phi,\D
  \phi, \fourg, \targetg]$. The stress energy tensor is of the form
\begin{equation}
T = P_{AB} \, \D \phi^{A} \otimes \D \phi^{B} + P \, \fourg ,
\label{CIR-5}
\end{equation}
where the functions $P_{AB}$ and $P$ may depend on $\phi$, $\D \phi$,
the spacetime metric $\fourg$ and the target metric $\targetg$. If
$\phi$ is invariant under the action of a Killing field $\xi$ -- in
the sense that $L_{\xi} \phi^{A} = 0$ for each component $\phi^{A}$ of
$\phi$ -- then the one-form $T(\xi)$ becomes proportional to $\xi$:
$T(\xi) = P \, \xi$. By virtue of the Killing field
identity~(\ref{CIR-1}), this implies that the twist of $\xi$ is
closed. Hence, the staticity and the circularity issue for
self-gravitating scalar mappings can be established, under appropriate
conditions, as in the vacuum case. From this one concludes that
(strictly) stationary non-rotating black-hole configuration of
self-gravitating scalar fields are static if $L_{k} \phi^{A} = 0$,
while stationary and axisymmetric ones are circular if $L_{k} \phi^{A}
= L_{m} \phi^{A} = 0$.

\subsection{Two-dimensional elliptic equations}
\label{subsec-BVCM}

The vacuum and the EM equations in the presence of a Killing symmetry
describe harmonic maps into coset manifolds, coupled to
three-dimensional gravity (see Section~\ref{sec-SST}). This feature is
shared by a variety of other self-gravitating theories with scalar
(moduli) and Abelian vector fields (see Section~\ref{subsec-MF}), for
which the field equations assume the form~(\ref{SEMS-9}):
\begin{equation}
\barR_{ij} = \frac{1}{4} \,
\trMH{{\cal J}_{i} \, {\cal J}_{j}} \,,
\quad \D \barast {\cal J} = 0 \,,
\label{BVCM-1}
\end{equation}
The current one-form ${\cal J}= \Phi^{-1}\D \Phi$
is given in terms of the Hermitian matrix $\Phi$,
which comprises the norm and the
generalized twist potential of the Killing field,
the fundamental scalar fields and the electric and magnetic
potentials arising on the effective level for each
Abelian vector field.
If the dimensional reduction is performed with respect
to the axial Killing field $m = \partial_{\varphi}$ with
norm $e^{-2\lambda} := \sprod{m}{m}$, then $\bar{R}_{ij}$ is the
Ricci tensor of the
pseudo-Riemannian three-metric $\barg$, defined by
\begin{equation}
\fourg = e^{-2\lambda} (\D \varphi + a)^{2} +
e^{2\lambda} \, \barg .
\label{BCVM-2}
\end{equation}

In the stationary and axisymmetric case under consideration, there
exists, in addition to $m$, an asymptotically--time-like Killing field
$k$. Since $k$ and $m$ fulfill the orthogonal-integrability conditions,
the spacetime metric can locally be written in a
(2+2)-block diagonal form. Hence, the
circularity property implies that
\begin{itemize}
\item
$(\Sigma,\barg)$ is a \textit{static} pseudo-Riemannian
three-dimensional manifold with metric
$\barg = -\rho^{2} \D t^{2} + \gtiltens$;
\item
the connection $a$ is orthogonal to the two-dimensional
Riemannian manifold $(\tilde{\Sigma},\gtiltens)$, that is,
$a = a_{t} \, \D t$;
\item
the functions $a_{t}$ and $\tilde{g}_{ab}$ do not depend
on the coordinates $t$ and $\varphi$.
\end{itemize}
With respect to the resulting Papapetrou metric~\cite{P53G},
\begin{equation}
\fourg = e^{-2\lambda} (\D \varphi + a_{t} \, \D t)^{2} + e^{2\lambda}
\left( - \rho^{2} \D t^{2} + \gtiltens \right) ,
\label{BCVM-3}
\end{equation}
the field equations~(\ref{BVCM-1}) become
a set of partial differential equations on the two-dimensional
Riemannian manifold $(\tilde{\Sigma},\gtiltens)$:
\begin{equation}
\Delta_{\scriptsize\gtiltens} \rho = 0 \, ,
\label{BCVM-3a}
\end{equation}
\begin{equation}
\tilde{R}_{ab} - \frac{1}{\rho}
\tilde{\nabla}_{b} \tilde{\nabla}_{a} \rho = \frac{1}{4}
\, \trMH{{\cal J}_{a} \, {\cal J}_{b}} \,,
\label{BCVM-3b}
\end{equation}
\begin{equation}
\tilde{\nabla}^{a} \left( \rho \, J_{a} \right) = 0 \,,
\label{BCVM-3c}
\end{equation}
as is seen from the standard reduction of the Ricci tensor
$\bar{R}_{ij}$ with respect to the static three-metric
$\barg = -\rho^{2} \D t^{2} + \gtiltens$.
Further ${\cal J}_{t} = 0$ and
$\barast {\cal J}= -\rho \, \D t \wedge \tilde{\ast} {\cal J}$.

The last simplification of the field equations is obtained by choosing
$\rho$ as one of the coordinates on
$(\tilde{\Sigma},\gtiltens)$. Roughly speaking, this follows from the
fact that
$\rho^2:=\fourg_{t\varphi}^2-\fourg_{tt}\fourg_{\varphi\varphi}$ is
non-negative, that its square root $\rho$ is harmonic (with respect to
the Riemannian two-metric $\gtiltens$), and that the domain of outer
communications of a stationary black-hole spacetime is simply
connected; see~\cite{ChNguyen,ChCo,ChUone} for details. The function
$\rho$ and the conjugate harmonic function $z$ are called Weyl
coordinates. With respect to these, the metric $\gtiltens$ becomes
manifestly conformally flat, and one ends up with the spacetime metric
\begin{equation}
\fourg = -\rho^{2} e^{-2\lambda} \D t^{2}+
e^{-2\lambda} \left( \D \varphi + a_{t} \D t \right)^{2} + e^{2\lambda}
\, e^{2\/ h} \left( \D \rho^{2} + \D z^{2} \right),
\label{BCVM-4}
\end{equation}
the $\sigma$-model equations
\begin{equation}
\partial_{\rho} \left( \rho \, {\cal J}_{\rho} \right) +
\partial_{z} \left( \rho \, {\cal J}_{z} \right) = 0 ,
\label{BCVM-5}
\end{equation}
and the remaining Einstein equations
\begin{equation}
\partial_{\rho}h = \frac{\rho}{8} \,
\trMH{{\cal J}_{\rho} {\cal J}_{\rho} - {\cal J}_{z} {\cal J}_{z}}\,,
\quad
\partial_{z}h = \frac{\rho}{4} \,
\trMH{{\cal J}_{\rho} {\cal J}_{z}} \,,
\label{BCVM-6}
\end{equation}
for the function $h(\rho,z)$. It is not hard to verify that
Eq.~(\ref{BCVM-5}) is the integrability condition for
Eqs.~(\ref{BCVM-6}). Since Eq.~(\ref{BCVM-3b}) is conformally
invariant, the metric function $h(\rho,z)$ does not appear in the
$\sigma$-model equation~(\ref{BCVM-5}). Taking into account that
$\rho$ is non-negative, the stationary and axisymmetric equations
reduce to an elliptic system for a matrix $\Phi$ on a flat
half-plane. Once the solution to Eq.~(\ref{BCVM-5}) is known, the
remaining metric function $h(\rho,z)$ is obtained from
Eqs.~(\ref{BCVM-6}) by quadrature.

\subsection{The Ernst equations}
\label{subsec-EE}

The circular $\sigma$-model equations~(\ref{BCVM-5}) for the EM
system, with target space $SU(2,1)/S(U(2) \times U(1))$, are called
\emph{Ernst equations}. Here, again, we consider the dimensional
reduction with respect to the \textit{axial} Killing field. The fields
can be parameterized in terms of the Ernst potentials $\Lambda = -
\phi + i \psi$ and $\erpot = - e^{-2\lambda} - \Lambda \bar{\Lambda} +
i Y$, where the four scalar potentials are obtained from
Eqs.~(\ref{SEMS-3}) and~(\ref{SEMS-4}) with $\xi = m$. Instead of
writing out the components of Eq.~(\ref{BCVM-5}) in terms of $\Lambda$
and $\erpot$, it is more convenient to consider Eqs.~(\ref{SEMS-6}),
and to reduce them with respect to a static metric $\barg = -\rho^{2}
\D t^{2} + \gtiltens$ (see Section~\ref{subsec-BVCM}). Introducing the
complex potentials $\epspot$ and $\lambda$ according to
\begin{equation}
\epspot = \frac{1-\erpot}{1+\erpot} , \quad
\lambda = \frac{2 \, \Lambda}{1+\erpot} ,
\label{EE-1}
\end{equation}
one easily finds the two equations
\begin{equation}
\tDelta \zeta +
\barsprod{\D \zeta}{
\frac{\D \rho}{\rho} +
\frac{2 \,( \bar{\epspot} \D \epspot + \bar{\lambda} \D \lambda)}
{1 - |\epspot|^{2} - |\lambda|^{2}}} = 0 ,
\label{EE-2}
\end{equation}
where $\zeta$ stands for either of the complex potentials $\epspot$ or
$\lambda $. Here we have exploited the conformal invariance of the
equations and used both the Laplacian $\tDelta$ and the inner product
with respect to a flat two-dimensional metric $\delta$. Indeed,
consider two black-hole solutions, then each black hole comes with its
own metric $\gtiltens$. However, the equation is conformally
covariant, and the $(\rho, z)$ representation of the metric is
manifestly conformally flat, with the same domain of coordinates for
both black holes. This allows one to view the problem as that of two
different Ernst maps defined on the same flat half-plane in
$(\rho,z)$-coordinates.

\subsubsection{A derivation of the Kerr--Newman metric}

The Kerr--Newman metric is easily derived within this formalism. For
this it is convenient to introduce, first, prolate spheroidal
coordinates $x$ and $y$, defined in terms of the Weyl coordinates
$\rho$ and $z$ by
\begin{equation}
\rho^{2} = \mu^{2} \left(x^{2}-1)(1-y^{2} \right) \, ,
\quad
z = \mu \, x \/ y \, ,
\label{EE-3}
\end{equation}
where $\mu$ is a constant. The domain of outer communications, that
is, the upper half-plane $\rho \geq 0$, corresponds to the semi-strip
${\cal S} = \{(x,y) | x \geq 1\, , |y| \leq 1 \}$. The boundary $\rho
= 0$ consists of the horizon ($x = 0$) and the northern ($y=1$) and
southern ($y = -1$) segments of the rotation axis. In terms of $x$ and
$y$, the metric $\gtiltens$ becomes $(x^2 - 1)^{-1} \D x^2 + (1 -
y^2)^{-1} \D y^2$, up to a conformal factor, which does not enter
Eqs.~(\ref{EE-2}). The Ernst equations finally assume the form
($\epspot_{x}: = \partial_{x} \epspot$, etc.)
\begin{eqnarray}
\lefteqn{
\left( 1 - |\epspot|^{2} - |\lambda|^{2} \right)
\left\{
\partial_{x} (x^{2}-1) \partial_{x} +
\partial_{y} (1-y^{2}) \partial_{y}
\right\} \zeta =
} \nonumber \\
&& - 2 \left\{ (x^{2}-1)
\left(\bar{\epspot} \epspot_{x} +
\bar{\lambda} \lambda_{x} \right) \partial_{x} +
(1-y^{2})
\left(\bar{\epspot} \epspot_{y} +
\bar{\lambda} \lambda_{y} \right) \partial_{y}
\right\} \zeta \,,
\label{EE-4}
\end{eqnarray}
where $\zeta$ stand for $\epspot$ or $\lambda$.
A particularly simple solution to those equations is
\begin{equation}
\epspot = p \/ x + i \, q \/ y \,, \quad
\lambda = \lambda_{0} \,, \quad
\mbox{where} \; p^{2} + q^{2} + \lambda_{0}^{2} = 1 \,,
\label{EE-5}
\end{equation}
with real constants $p$, $q$ and $\lambda_{0}$.
The norm $e^{-2\lambda}$, the twist potential $Y$
and the electro-magnetic potentials
$\phi$ and $\psi$ (all defined with respect to the axial Killing
field) are obtained from the above solution by using
Eqs.~(\ref{EE-1}) and the expressions
$e^{-2\lambda}=-\mbox{Re}(\erpot)-|\Lambda|^{2}$,
$Y=\mbox{Im}(\erpot)$, $\phi = -\mbox{Re}(\Lambda)$,
$\psi = \mbox{Im}(\Lambda)$. The off-diagonal element
of the metric, $a = a_{t} \D t$, is obtained by
integrating the twist expression~(\ref{REHA-3}), where the
twist one-form is given in Eq.~(\ref{SEMS-4}), and the
Hodge dual in Eq.~(\ref{REHA-3}) now refers to the
decomposition~(\ref{BCVM-2}) with respect to the axial Killing field.
Eventually, the metric function $h$ is obtained from
Eqs.~(\ref{BCVM-6}) by quadratures.

The solution derived so far is the ``conjugate''
of the Kerr--Newman solution~\cite{SC91BK}. In order to
obtain the Kerr--Newman metric itself, one has to perform
a rotation in the $t\varphi$-plane: The spacetime metric
is invariant under $t \rightarrow \varphi$,
$\varphi \rightarrow -t$, if $e^{-2\lambda}$, $a_{t}$ and $e^{2h}$ are
replaced by ${\alpha}e^{-2\lambda}$, ${\alpha}^{-1}a_{t}$ and ${\alpha}e^{2h}$, where
${\alpha} := a_{t}^2 - e^{4\lambda} \rho^{2}$. This
additional step in the derivation of the Kerr--Newman metric
is necessary because the Ernst potentials were defined
with respect to the axial Killing field $\partial_{\varphi}$.
If, on the other hand,
one uses the stationary Killing field $\partial_{t}$, then the
Ernst equations are singular at the boundary of the ergoregion.

In terms of Boyer--Lindquist coordinates,
\begin{equation}
r = m \, (1 + px) , \quad \cos \! \vartheta = y ,
\label{EE-6}
\end{equation}
one eventually finds the Kerr--Newman metric in the familiar form:
\begin{equation}
\fourg = - \frac{\Delta}{\Xi} \left[
\D t - \alpha \sin^2 \! \vartheta \D \varphi \right]^{2} +
\frac{\sin^2 \! \vartheta}{\Xi} \left[
(r^2 + \alpha^2) \D \varphi - \alpha \D t \right]^{2} +
\Xi \left[ \frac{1}{\Delta} \D r^2 + \D \vartheta^2 \right] ,
\label{EE-8}
\end{equation}
where the constant $\alpha$ is defined by
$a_{t} := \alpha \sin^2 \! \vartheta$. The expressions
for $\Delta$, $\Xi$ and the electro-magnetic vector potential $A$
show that the Kerr--Newman solution is characterized by the
total mass $M$, the electric charge $Q$, and the angular
momentum $J = \alpha M$:
\begin{equation}
\Delta = r^2 - 2 M r + \alpha^2 + Q^2 , \quad
\Xi = r^2 + \alpha^2 \cos^2 \! \vartheta .
\label{EE-9}
\end{equation}
\begin{equation}
A = \frac{Q}{\Xi} \, r \,
\left[ \D t - \alpha \sin^2 \! \vartheta \D \varphi \right] .
\label{Ahatt}
\end{equation}

\subsection{The uniqueness theorem for the Kerr--Newman solution}
\label{subsec-UTKN}

In order to establish uniqueness of the Kerr--Newman
metric among the stationary and axisymmetric
black-hole configurations, one has to show that
two solutions of the Ernst equations~(\ref{EE-5})
are equal if they are subject to black-hole boundary conditions on
%
\newcommand{\myS}{\mathbb{{\cal S}}}
$\partial {\myS}$, where ${\myS}$ is the
half-plane ${\myS} = \{(\rho,z) | \rho\geq0 \}$.  
Carter proved non-existence of linearized vacuum perturbations near
Kerr by means of a divergence identity~\cite{BC71bd}, which Robinson
generalized to electrovacuum spacetimes~\cite{DCR74}.

\subsubsection{Divergence identities}
\label{subsec-DivId}

Considering two \textit{arbitrary} solutions of the Ernst equations,
Robinson was able to construct an identity~\cite{DCR75Ident}, the
integration of which proved the uniqueness of the Kerr metric. The
complicated nature of the Robinson identity dashed the hope of finding
the corresponding electrovacuum identity by trial and error methods
(see, e.g.,~\cite{BC79GT}). The problem was eventually solved when
Mazur~\cite{PM82UT, PM84b} and Bunting~\cite{GB83UT} independently
derived divergence identities useful for the problem at
hand. Bunting's approach, applying to a general class of harmonic
mappings between Riemannian manifolds, yields an identity, which
enables one to establish the uniqueness of a harmonic map if the
target manifold has negative curvature. {We refer the reader to
  Sections~\ref{ss11VI11.2} and~\ref{ss14VI11.1} (see
  also~\cite{BC85BM}) for discussions related to Bunting's method.}

So, consider two solutions of the Ernst equations associated to, a
priori, distinct black-hole spacetimes, each endowed with its own
metric. As discussed in Section~\ref{subsec-EE}, Weyl coordinates and
conformal invariance allow us to view the Ernst equations as equations
on a flat half-plane; alternatively, they may be seen as equations for
an axisymmetric field on three-dimensional flat space. The Mazur
identity~(\ref{TMI-2}) applies to the relative difference $\Psi =
\Phi_{2} \Phi_{1}^{-1} - \mathbb{1}$ of the associated Hermitian
matrices and implies (see Section~\ref{subsec-TMI} for details and
references)
\begin{equation}
\trMH{ \Delta_{\gamma} \Psi } =
\mbox{Trace} \barsprod{{\cal M}}{{\cal M}^{\dagger}} ,
\label{UTKN-1}
\end{equation}
where $\Delta_{\gamma}$ is the Laplace--Beltrami operator of the flat
metric $\gamma=d\rho^2+dz^2+\rho^2 d\varphi^2$; also recall that
${\cal M} = g_{1}^{-1} {\cal J}^{\dagger}_{\triangle} g_{2}$, with
${\cal J}^{\dagger}_{\triangle}$ the difference between the currents.

The reduction of the EM equations with respect to the axial Killing
field yields $\sigma$-model equations with $SU(2,1)/S(U(2) \times
U(1))$ target (see Section~\ref{subsec-SEMS}),  in vacuum reduces
to $SU(2)/S(U(1) \times U(1))$ (see Section~\ref{subsec-TSVG}). Hence,
the above formula applies to both the stationary and axisymmetric
vacuum or electrovacuum field equations. Now, relying on axisymmetry
once more, we can reduce the previous Mazur identity to an equation on
the flat half-plane $({\myS}, \delta)$; integrating and using Stokes'
theorem leads to
\begin{equation}
\int_{\partial{\myS}} \rho \, {\ast} \,
\trMH{ \D \Psi } =
\int_{{\myS}} \rho \,
\mbox{Trace} \barsprod{{\cal M}}{{\cal M}^{\dagger}}
\, {\eta}_{\delta},
\label{UTKN-2}
\end{equation}
where the volume form ${\eta}_{\delta}$ and the Hogde dual ${\ast}$
are related to the flat metric $\delta=d\rho^2+dz^2$.

The uniqueness of the Kerr--Newman metric should follow now from
\begin{itemize}
\item the fact that the integrand on the right-hand side is
  non-negative, and

\item the fact that the boundary at infinity on the left-hand side
  vanishes for two solutions with the same mass, electric charge and
  angular momentum, and

\item the expectation that the integral over the axis and horizons,
  where the integrand becomes singular, vanishes for black-hole
  configurations with the same quotient-space structure.
\end{itemize}

In order to establish that $\rho \, \trMH{\D \Psi} = 0$ on the
boundary $\partial {\myS}$ of the half-plane,%
\epubtkFootnote{A workable formula for $\mbox{Trace}\Psi$ is provided
 by~\eqref{sigma} (compare~\cite[eq. 4.14]{PM82UT}).}
one needs the asymptotic behavior and the boundary and regularity
conditions of all potentials. One expects that $\rho \, \trMH{\D
  \Psi}$ vanishes on the horizon, the axis and at infinity, provided
that the solutions have the same mass, charge and angular momentum,
but no complete analysis of this has been presented in the literature;
see~\cite{GW90Rot} for some partial results. Fortunately, the
supplementary difficulties arising from the need to control the
derivatives of the fields disappear when the distance-function
approach described in the next Section~\ref{ss14VI11.1} is used.

\subsubsection{The distance function argument}
\label{ss14VI11.1}

An alternative to the divergence identities above is provided by the
observation that the distance $d(\phi_1,\phi_2)$ between two harmonic
maps $\phi_a$, $a=1,2$, with negatively curved target manifold is
subharmonic~\cite[Lemma~8.7.3 and Corollary~8.6.4]{MR99g:53025} (see
also the proof of Lemma~2 in~\cite{Weinstein:Hadamard} following
results in~\cite{SY:Compact}):
\begin{equation}
\label{Deltad}
  \Delta_\delta d(\phi_1,\phi_2)\ge 0\;;
\end{equation}
compare~\eqref{sigmaSubHarm}. Here $d$ is the distance function
between points on the target manifold and $\Delta_\delta$ the flat
Laplacian on $\R^3$. It turns out that the Ernst equations for the
Einstein--Maxwell equations fall in this category; in the vacuum case
this is obvious, as the target space is then the two-dimensional
hyperbolic space. This is somewhat less evident for the
Einstein--Maxwell Ernst map, and can be checked by a direct
calculation, or can be justified by general considerations about
symmetric spaces; more precisely this follows from~\cite[Theorem
 3.1]{Helgason} after noting that the target spaces of the maps under
consideration are of non-compact type (see also~\cite{WeinsteinDuke}).

Using this observation, the key to uniqueness is provided by the
following non-standard version of the maximum principle:

\begin{Proposition}{\rm{\cite[Appendix~C]{CLW}}}
\label{Pvn}
Let $\mcA$ denote the $z$-axis in $\R^3$, and let $f\in
C^0(\R^3\setminus \mcA)$ satisfy
\begin{equation}
\Delta_\delta f\ge 0\qquad\mbox{in}\ \R^3\setminus \mcA,
\quad \mbox{in the distributional sense},
\label{1a}
\end{equation}
\begin{equation}
0\le f\le 1,\qquad \mbox{on}\ \R^3\setminus \mcA,
\label{2a}
\end{equation}
and
\begin{equation}
\lim_{(x,y,z)\in \R^3\setminus \mcA, |(x,y,z)|\to \infty} f(x,y,z)=0.
\label{3a}
\end{equation}
Then
$$
f\equiv 0, \qquad \mbox{on}\ \R^3\setminus \mcA.
$$
\end{Proposition}

Hence, to prove uniqueness it remains to verify that
$d(\phi_1,\phi_2)$ is bounded on $\R^3\setminus\mcA$, and that
$$
 f(x):= \frac{d(\phi_1(x),\phi_2(x))}{\sup_y d(\phi_1(y),\phi_2(y))}
$$
goes to zero at infinity. The latter property follows immediately from
asymptotic flatness. The main work is thus to prove that $f$ remains
bounded near the axis. Here one needs to keep in mind that the
$(\rho,z)$ coordinate system is constructed in an implicit way by PDE
techniques, and that the whole axis is singular from the PDE point of
view because of factors of $\rho$ and $\rho^{-1}$ in the
equations. In particular the associated harmonic maps tend to infinity
in the target manifold when the axis of rotation $\mcA$ is
approached. So the proof of boundedness of $f$ requires considerable
effort, with the first complete analysis for non-degenerate horizons
in~\cite{ChCo}. The major challenge are points where the axes of
rotation meet the horizons. The degenerate horizons, first settled
in~\cite{ChNguyen}, provide supplementary difficulties. The proof of
boundedness of $f$ near degenerate horizons proceeds via H{\'{a}}j{\'{\i}}{\v{c}}ek's
Theorem~\cite{Hajicek3Remarks} (rediscovered independently by
Lewandowski and Paw\l owski~\cite{LP1}, see
also~\cite{KunduriLucietti2}), that the near-horizon geometry of
degenerate axisymmetric Killing horizons with spherical cross-sections
coincides with that of the Kerr--Newman solutions.

\newpage

\section{Acknowledgments}
\label{ack}

MH takes pleasure in thanking Othmar Brodbeck, Gary Gibbons, Domenico
Giulini, Dieter Maison, Gernot Neugebauer, Norbert Straumann, Michael
Volkov and Robert Wald for helpful discussions. PTC acknowledges
bibliographical advice from Michael Volkov.

This work was supported by SNSF Grant P-21-41840.94 (MH), by a Polish
Ministry of Science and Higher Education grant Nr N N201 372736 grant
(PTC), by projects PTDC/MAT/108921/2008 and CERN/FP/116377/2010, and
by CAMSDG, through FCT plurianual funding (JLC).


\newpage

\bibliography{refs} 

\end{document}